\pdfoutput=1

\documentclass[twocolumn]{emulateapj}

\usepackage{amsmath}
\usepackage{graphicx}
\usepackage{color}
\usepackage[colorlinks]{hyperref}
\hypersetup{
    colorlinks,	
    citecolor=blue,
}

\newcommand{\be}{\begin{eqnarray}}
\newcommand{\ee}{\end{eqnarray}}

\def\jcap{JCAP}

\shorttitle{Impact of the disk thickness on reflection spectra}
\shortauthors{Tripathi et al.}

\begin{document}

\title{Impact of the disk thickness on X-ray reflection spectroscopy measurements}

\author{Ashutosh~Tripathi\altaffilmark{1}, Askar~B.~Abdikamalov\altaffilmark{1,2}, Dimitry~Ayzenberg\altaffilmark{3}, Cosimo~Bambi\altaffilmark{1,\dag} and Honghui~Liu\altaffilmark{1}}

\altaffiltext{1}{Center for Field Theory and Particle Physics and Department of Physics, 
Fudan University, 200438 Shanghai, China. \email[\dag E-mail: ]{bambi@fudan.edu.cn}}
\altaffiltext{2}{Ulugh Beg Astronomical Institute, Tashkent 100052, Uzbekistan}
\altaffiltext{3}{Theoretical Astrophysics, Eberhard-Karls Universit\"at T\"ubingen, D-72076 T\"ubingen, Germany}

\begin{abstract}
In a previous paper, we presented an extension of our reflection model {\tt relxill\_nk} to include the finite thickness of the accretion disk following the prescription in \citet{2018ApJ...855..120T}. In this paper, we apply our model to fit the 2013 simultaneous observations by \textsl{NuSTAR} and \textsl{XMM-Newton} of the supermassive black hole in MCG--06--30--15 and the 2019 \textsl{NuSTAR} observation of the Galactic black hole in EXO~1846--031. The high-quality data of these spectra had previously led to precise black hole spin measurements and very stringent constraints on possible deviations from the Kerr metric. We find that the disk thickness does not change previous spin results found with a model employing an infinitesimally thin disk, which confirms the robustness of spin measurements in high radiative efficiency disks, where the impact of disk thickness is minimal. Similar analysis on lower accretion rate systems will be an important test for measuring the effect of disk thickness on black hole spin measurements.
\end{abstract}


\section{Introduction}

Relativistic reflection features are common in the X-ray spectra of Galactic black holes and AGN~\citep{1989MNRAS.238..729F,1995Natur.375..659T,2007MNRAS.382..194N}. They are thought to be produced from illumination of the inner part of an accretion disk by a hot corona~\citep{1995MNRAS.277L..11F,2013Natur.494..449R}. In the rest-frame of the gas in the disk, the reflection spectrum is characterized by a soft excess below 2~keV, narrow fluorescent emission lines in the 1-10~keV band, in particular the iron K$\alpha$ complex at 6.4~keV for neutral iron and up to 6.97~keV for H-like iron ions, and a Compton hump peaked at 20-30~keV~\citep{1995MNRAS.273..837M,2005MNRAS.358..211R,2010ApJ...718..695G}. The reflection spectrum of the whole disk detected far from the source is the result of the sum of the reflection emission at different points on the disk and relativistic effects occurring in the strong gravity region near the black hole (gravitational redshift, Doppler boosting, light bending)~\citep{1989MNRAS.238..729F,1991ApJ...376...90L}. The analysis of the relativistic reflection features in the spectrum of a source can thus be used to study the accretion process in the inner part of the accretion disk, measure black hole spins, and even test fundamental physics~\citep{2013mams.book.....B,2014SSRv..183..277R,2017RvMP...89b5001B}.

The last decade has seen tremendous progress in the analysis of these reflection features in the spectra of accreting black holes, thanks to new observational facilities and more sophisticated theoretical models. However, all the available theoretical models still have a number of simplifications~\citep{2020arXiv201104792B}, so caution is necessary in any attempt to infer precision measurements of the properties of these sources. Moreover, the next generation of X-ray observatories, such as \textsl{eXTP}~\citep{2016SPIE.9905E..1QZ} or \textsl{Athena}~\citep{2013arXiv1306.2307N}, promise to provide unprecedented high-quality data, and more sophisticated synthetic reflection spectra will be required to fit their data. There are thus important efforts among the community to further develop the current reflection models.

In the past 5~years, our group has developed the model {\tt relxill\_nk}~\citep{2017ApJ...842...76B,2019ApJ...878...91A}, which is an extension of the {\tt relxill} package~\citep{2013ApJ...768..146G,2014ApJ...782...76G} to non-Kerr spacetimes. The key-feature of {\tt relxill\_nk} is that it does not employ the Kerr solution of the Einstein Equations as the background metric. Synthetic reflection spectra are instead calculated in a more general background that includes the Kerr metric as a special case. The spacetime is characterized by some ``deformation parameters''. The latter vanish for the Kerr metric and can thus be used to quantify possible deviations from the Kerr geometry. From the comparison of reflection-dominated X-ray data with synthetic reflection spectra, we can constrain the values of these deformation parameters and thus test the Kerr black hole hypothesis \citep{2018PhRvL.120e1101C}. In most of our past studies, we have used the parametric black hole spacetime suggested in \citet{2013PhRvD..88d4002J}, which is not a solution of a specific theory of gravity but an {\it ad hoc} deformation of the Kerr metric for these kinds of studies. However, the model can also employ theoretically motivated black hole solutions~\citep{2018PhRvD..98b4007Z,2020EPJC...80..622Z,2020arXiv200512958Z}.

{\tt relxill\_nk} has been used to analyze a number of reflection dominated spectra of Galactic black holes and AGN, and in some cases we were able to derive much stronger constraints than those found with other electromagnetic and gravitational wave techniques~\citep{2019ApJ...884..147Z,2019ApJ...875...56T,2020arXiv201013474T,2020arXiv201210669T}. However, more and more precise measurements necessarily need to be more and more accurate, and this requires continuing the development of the model~\citep{2020ApJ...899...80A,2021arXiv210110100A,2020arXiv201207469R}.

In \citet{2020ApJ...899...80A}, we implemented in {\tt relxill\_nk} an accretion disk of finite thickness following the prescription proposed in \citet{2018ApJ...855..120T}. We still have a Novikov-Thorne accretion disk perpendicular to the black hole spin and with the inner edge at the innermost stable circular orbit (ISCO), but -- unlike all the currently public relativistic reflection models assuming an infinitesimally thin disk -- the disk has a finite thickness, which increases as the mass accretion rate onto the black hole increases. In the present paper, we apply that model with a disk of finite thickness to fit the 2013 \textsl{NuSTAR}+\textsl{XMM-Newton} data of the supermassive black hole in MCG--06--30--15 and the 2019 \textsl{NuSTAR} data of the Galactic black hole in EXO~1846--031. These are high-quality data and have already been analyzed with the standard {\tt relxill\_nk} with an infinitesimally thin disk, providing among the most precise and accurate constraints of the deformation parameters of the Johannsen metric~\citep{2019ApJ...875...56T,2020arXiv201210669T}. Moreover, MCG--06--30--15 is a supermassive black hole observed from a low viewing angle, while EXO~1846--031 is a stellar-mass black hole observed from a very high inclination angle. They are thus two data sets particularly suitable for testing the development of our model.

The paper is organized as follows. In Section~\ref{s-disk}, we briefly review our reflection model with an accretion disk of finite thickness. In Section~\ref{s-mcg}, we analyze with our model the 2013 \textsl{NuSTAR}+\textsl{XMM-Newton} observations of the supermassive black hole in MCG--06--30--15. In Section~\ref{s-exo}, we fit the 2019 \textsl{NuSTAR} data of the Galactic black hole in EXO~1846--031. We discuss our results in Section~\ref{s-dis}.


\section{Accretion disk of finite thickness} \label{s-disk}

\begin{figure*}
\begin{center}
\includegraphics[width=18.0cm,trim={0cm 0cm 0cm 0cm},clip]{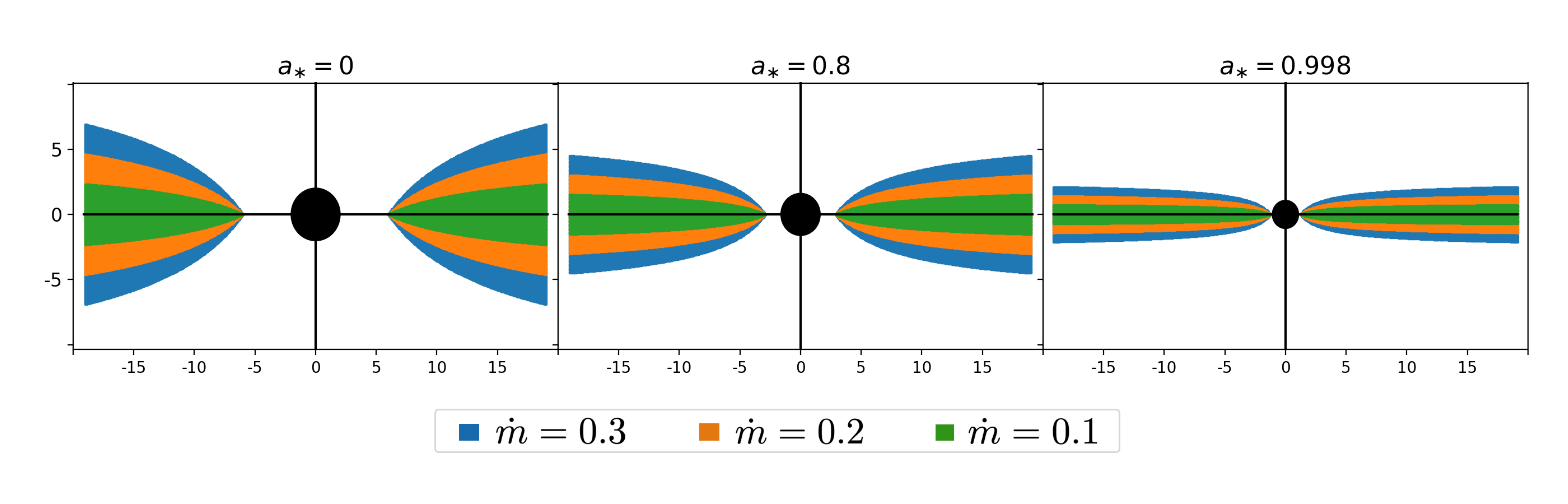}
\end{center}
\vspace{-0.7cm}
\caption{Profiles of accretion disks for the black hole spin parameter $a_* = 0$, 0.8, and 0.998 and the Eddington-scaled mass accretion rate 
$\dot{m} = 0.1$, 0.2, and 0.3. Figure following \citet{2018ApJ...868..109T}. \label{fig:disk} }
\end{figure*}

In \citet{2020ApJ...899...80A}, we extended our reflection model {\tt relxill\_nk} to include the accretion disk geometry proposed in~\citet{2018ApJ...855..120T}. Here we briefly review the set-up. We consider a geometrically thin and optically thick accretion disk with the mid-plane orthogonal to the black hole spin angular momentum. The disk is radiation dominated and its pressure scale height is given by~\citep{1973A&A....24..337S}
	\be\label{eq:scale_height}
	H=\frac{3}{2}\frac{\dot{m}}{\eta} \left[1-\left(\frac{r_{\rm ISCO}}{\rho}\right)^{\frac{1}{2}}\right],
	\ee
	where $\rho=r\sin\theta$ is the pseudo-cylindrical radius, $r_{\rm ISCO}$ is the radius of the ISCO and in this paper it is always assumed to be the location of the inner edge of the accretion disk, $\dot{m} \equiv \dot{M}/\dot{M}_{\rm Edd}$ is the Eddington-scaled mass accretion rate, and $\eta$ is the radiative efficiency defined as $\eta=1 - E_{\rm ISCO}$ with $E_{\rm ISCO}$ the specific energy of a test particle at the ISCO. The surface of the accretion disk is determined by the function $z(\rho)=2H$ and all matter of the disk at a given pseudo-cylindrical radius $\rho$ has the same angular velocity. In Eq.~(\ref{eq:scale_height}), $\eta$ and $r_{\rm ISCO}$ depend on the spacetime metric. For a given spacetime metric, the Eddington-scaled mass accretion rate regulates the thickness of the accretion disk. Fig.~\ref{fig:disk} shows the impact of $\dot{m}$ in the Kerr spacetime for different values of the black hole spin parameter.

\begin{figure*}
\begin{center}
\includegraphics[width=18.0cm,trim={2.5cm 0cm 3cm 2cm},clip]{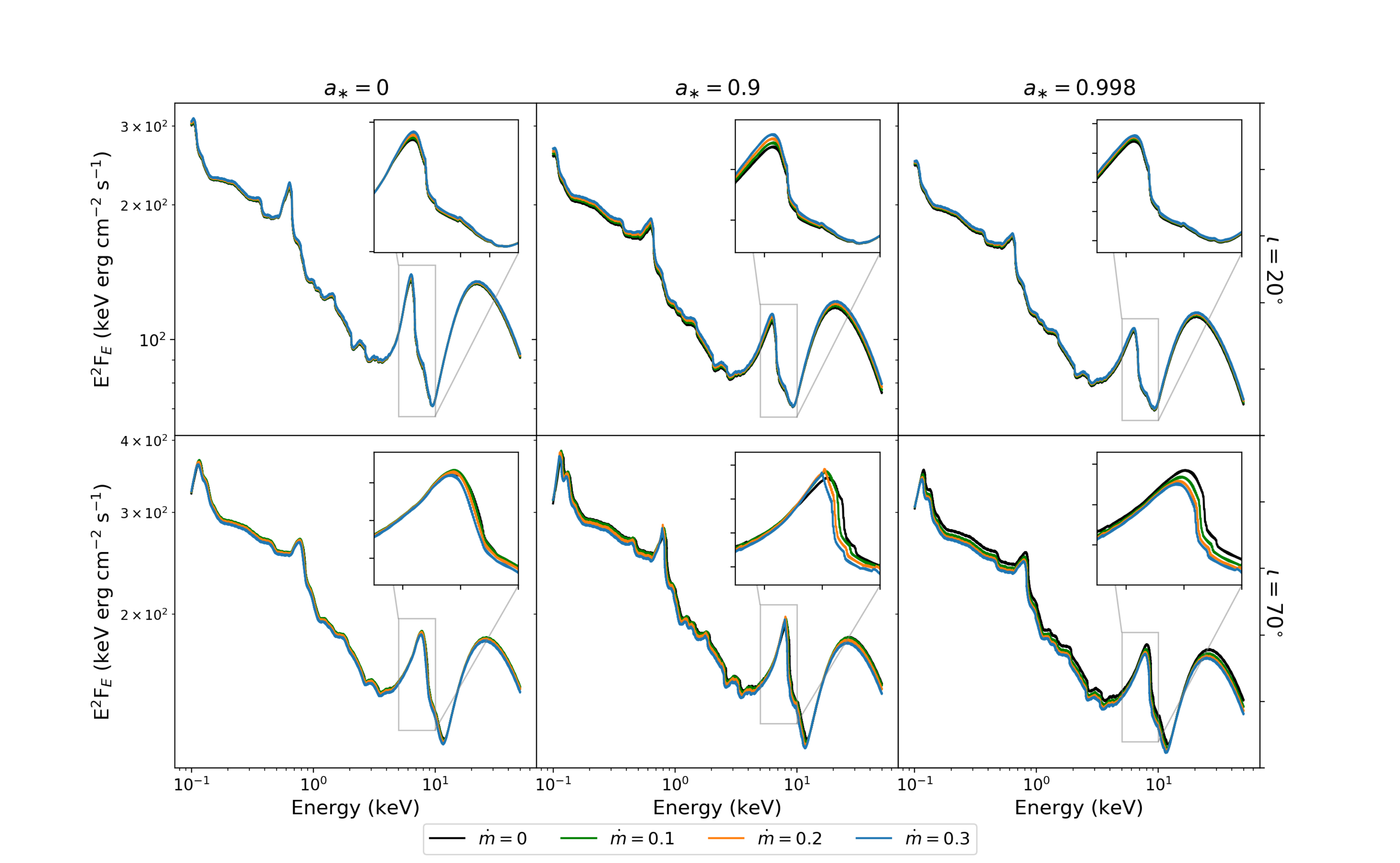}
\end{center}
\vspace{-0.2cm}
\caption{Synthetic disk reflection spectra in the Kerr spacetime for different values of the black hole spin ($a_* = 0$, 0.9, and 0.998), inclination angle of the disk ($\iota = 20^\circ$ and $70^\circ$), and Eddington-scaled mass accretion rate ($\dot{m} = 0$, 0.1, 0.2, and 0.3). The inner edge of the disk is set at the ISCO radius and the emissivity profile is described by a power-law with emissivity index $q=3$. The radiation illuminating the disk has the photon index $\Gamma = 2$ and the high-energy cutoff $E_{\rm cut} = 300$~keV. The iron abundance of the disk is $A_{\rm Fe} = 1$ (solar iron abundance) and its ionization is $\log\xi = 3.1$. \label{fig:eemod} }
\end{figure*}

In the model suite {\tt relxill\_nk}, relativistic effects and disk structure are included within the formalism of the transfer function~\citep{1975ApJ...202..788C,1995CoPhC..88..109S,2010MNRAS.409.1534D}. The observed spectrum can be written as
	\be\label{eq:flux1}
	F_{\rm o}(E_{\rm o}) = \int I_{\rm o}(E_{\rm o}, X, Y) d\tilde{\Omega},
	\ee
	where $I_{\rm o}(E_{\rm o}, X, Y)$ is the specific intensity of the radiation with energy $E_{\rm o}$ on the observer's screen with the Cartesian coordinates $X$ and $Y$, $d\tilde{\Omega}=dXdY/D^2$ is the element of the solid angle subtended by the disk image in the observer's screen, and $D$ is the distance of the source from the observer. Using the Liouville's theorem~\citep{1966AnPhy..37..487L}, we write $I_{\rm o}=g^3I_{\rm e}$, where $g=E_{\rm o}/E_{\rm e}$ is the redshift factor between the photon energy measured on the observer's screen, $E_{\rm o}$, and the photon energy at the emission point in the rest-frame of the gas in the disk, $E_{\rm e}$.

Eq.~(\ref{eq:flux1}) can be rewritten as
\be
F_{\rm o}(E_{\rm o})&=&\frac{1}{D^2}\int^{R_{\text{out}}}_{R_{\text{in}}}\int^{1}_{0}\frac{\pi r_{\rm e}g^{2}}{\sqrt{g^{*}(1-g^{*})}} \nonumber\\
&& \times f(g^{*},r_{\rm e},\iota) I_{\rm e}(E_{\rm e}, r_{\rm e},\theta_{\rm e})dg^{*}dr_{\rm e} \, , \label{eq:flux}
\ee
where $R_{\text{in}}$ and $R_{\text{out}}$ are, respectively, the inner and the outer edges of the accretion disk, $r_{\rm e}$ is the radial coordinate of the emission point in the accretion disk, $f$ is the transfer function defined as
\be
f(g^{*},r_{\rm e},\iota)=\frac{1}{\pi r_{\rm e}}g\sqrt{g^{*}(1-g^{*})}\left|\frac{\partial(X,Y)}{\partial(g^{*},r_{\rm e})}\right|,
\ee
$\left|\partial(X,Y)/\partial(g^{*},r_{\rm e})\right|$ is the Jacobian, $\theta_e$ is the emission angle in the rest frame of the gas, and $g^{*}$ is the relative redshift factor defined as
\be
g^{*}=\frac{g-g_{\text{min}}}{g_{\text{max}}-g_{\text{min}}} \in[0,1] \, ,
\ee
where $g_{\rm min}=g_{\rm min}(r_{\rm e}, \iota)$ and $g_{\rm max}=g_{\rm max}(r_{\rm e}, \iota)$ are the minimum and maximum values of the redshift factor $g$ for fixed values of emission radius $r_{\rm e}$ and the observer's viewing angle $\iota$.

With the formalism of the transfer function, we can easily separate the calculations of the spectrum at the emission point in the rest-frame of the gas in the disk and the effects related to the disk structure and the spacetime metric that modify such a spectrum and lead to that observer far from the source. This separation enables us to quickly calculate the observed reflection spectrum from the tabulated transfer functions without having to recalculate all photon trajectories, which is quite a time consuming step. Since the accretion disk has now a finite thickness, a region of the inner part of the accretion disk may not be visible to the distant observer. For the non-visible points of the accretion disk, the transfer function vanishes. The transfer function and the emission angles are calculated once and tabulated in a FITS (Flexible Image Transport System) table. During the data analysis process, the values of the transfer function corresponding to the set of input parameters are entered into Eq.~(\ref{eq:flux}) using interpolation schemes.

The table has a grid of three dimensions, namely the black hole spin parameter $a_{*}$, the deformation parameter\footnote{In this paper, we assume that the spacetime geometry is described by the Johannsen metric with the only non-vanishing deformation parameter $\alpha_{13}$~\citep[for more details, see][]{2013PhRvD..88d4002J}} $\alpha_{13}$, and the viewing angle $\iota$, which are sampled in a grid of $30 \times 30 \times 22$, respectively. The spin parameter grid is distributed in such a way that it becomes denser as the spin increases, as the change of the ISCO becomes faster at higher values of $a_{*}$. The grid of emission angles is evenly distributed in $0<\cos\iota<1$. The grid of the deformation parameter $\alpha_{13}$ is evenly distributed in the range $[-5,5]$ or the range imposed by constraints on $\alpha_{13}$, whichever is smaller.
For each set of grid points ($a_{*}$, $\alpha_{13}$, $\iota$), the accretion disk is divided into 100 emitting radii from $r_{\rm ISCO}$ to 1000~$r_{\rm g}$, where $r_{\rm g} = G_{\rm N}M/c^2$ is the gravitational radius of the black hole. For every emitting radius, the minimum and maximum redshift factors are calculated, as well as the transfer functions and the emission angles for 40 evenly spaced values of $g^{*}$ on each branch. The general relativistic ray tracing code, fully described in~\cite{2019ApJ...878...91A}, calculates the parameters required for the FITS file. The code solves the photon trajectories backwards in time from the observer's screen to the accretion disk. As we do not know {\it a priori} where every photon lands, an adaptive algorithm samples the initial conditions of the photon so that the photon lands at the desired value of $r_{\rm e}$ to calculate the redshift factor and the emission angle. The Jacobian and then the transfer function are calculated by firing two photons that land at the same emission radius $r_{\rm e}$ and two more adjacent photons landing at neighboring radii. Finally, a separate script processes all photon data and generates the FITS file for a specific value of $\dot{m}$. The current model structure does not allow the Eddington ratio to be a model parameter, as this would result in a too large FITS file. We thus have FITS tables for specific mass accretion rates (0\%, 5\%, 10\%, 20\%, and 30\%). Fig.~\ref{fig:eemod} shows our synthetic reflection spectra in the Kerr spacetime for different values of the black hole spin parameter $a_*$, inclination angle of the disk $\iota$, and mass accretion rate $\dot{m}$.


\section{MCG--06--30--15} \label{s-mcg}

The Narrow line Seyfert~1 galaxy MCG--06--30--15 is one of the most observed AGN in the X-ray band due to the presence of prominent reflection features. It is the source in which a broad iron line was observed for the first time~\citep{1995Natur.375..659T}. MCG--06--30--15 was observed by \textsl{ASCA}\citep{2002MNRAS.333..687S,2003PASJ...55..615M}, \textsl{RXTE} \citep{1999MNRAS.310..973L,2001ApJ...548..694V}, \textsl{BeppoSAX} \citep{1999A&A...341L..27G}, \textsl{XMM-Newton} \citep{2001MNRAS.328L..27W,2002MNRAS.335L...1F,2003MNRAS.340L..28F,2004MNRAS.348.1415V,2006ApJ...652.1028B}, \textsl{Suzaku} \citep{2007PASJ...59S.315M,2011PASJ...63..449N}. \citet{2008A&A...483..437M} and \citet{2011MNRAS.414.2345C} analyzed data from multiple instruments to better study the complexity of this source.

The X-ray spectrum of MCG--06--30--15 normally has a prominent and very broad iron line, so it is quite a good candidate for precision measurements from the analysis of its relativistic reflection features~\citep{2006ApJ...652.1028B}. \citet{2014ApJ...787...83M} analyzed simultaneous observations of \textsl{NuSTAR} and \textsl{XMM-Newton} and inferred the black hole spin parameter $a_* = 0.91_{-0.07}^{+0.06}$. The same data set was analyzed in \citet{2019ApJ...875...56T} with {\tt relxill\_nk} to test the Kerr nature of the compact object and we constrained the deformation parameters $\alpha_{13}$ and $\alpha_{22}$ of the Johannsen metric; we found $\alpha_{13} = 0.00_{-0.20}^{+0.07}$ (assuming $\alpha_{22} = 0$) and $\alpha_{22} = 0.0_{-0.1}^{+0.6}$ (assuming $\alpha_{13} = 0$), where the Kerr metric corresponds to the case $\alpha_{13} = \alpha_{22} = 0$. Spectral analyses of X-ray data from various missions confirm that the spectrum of MCG--06--30--15 is modified by the presence of warm absorbers~\citep{1996PASJ...48..211O,2000MNRAS.318..857L,2005ApJ...631..733Y}. Optical observations reveal the presence of a dusty neutral absorber~\citep{2003MNRAS.346..833T,2004MNRAS.353..319T}. Besides the complex spectrum, MCG--06--30--15 is also highly variable.

\begin{table*}
 \centering
 \renewcommand\arraystretch{1.5}
\begin{tabular}{ccccc}
\hline\hline
\hspace{0.1cm} Source \hspace{0.1cm} & \hspace{0.1cm} Mission \hspace{0.1cm} & \hspace{0.1cm} Observation ID \hspace{0.1cm} & \hspace{0.1cm} Observation Date \hspace{0.1cm} & \hspace{0.1cm} Exposure (ks) \hspace{0.1cm} \\
\hline\hline
MCG--06--30--15 & \textsl{NuSTAR} & 60001047002 & 2013-01-29 & 23 \\
&& 60001047003 & 2013-01-30 & 127 \\
&& 60001047005 & 2013-02-02 & 30 \\
& \textsl{XMM-Newton} & 0693781201 & 2013-01-29 & 134 \\
&& 0693781301 & 2013-01-31 & 134 \\
&& 0693781401 & 2013-02-02 & 49 \\
\hline
EXO~1846--031 & \textsl{NuSTAR} & 90501334002 & 2019-08-03 & 22 \\
\hline\hline
\end{tabular}
 \caption{\rm Summary of the observations analyzed in the present work. \label{t-obs}}
\end{table*}

\begin{figure}
\begin{center}
\includegraphics[width=8.5cm,trim={0.5cm 0.0cm 2.5cm 9.0cm},clip]{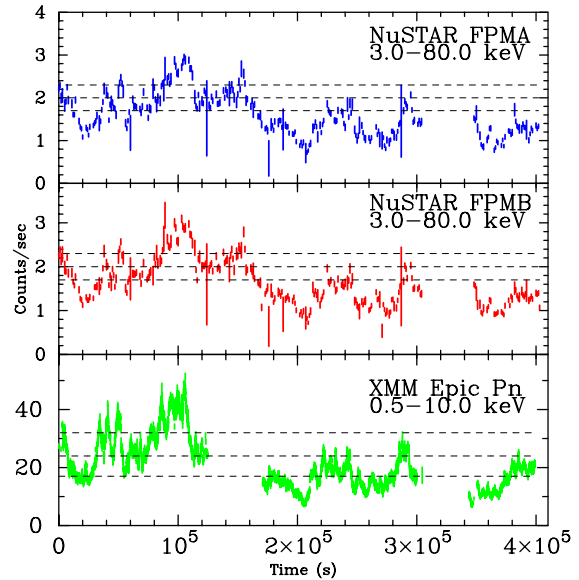}
\end{center}
\vspace{-0.6cm}
\caption{Light curves of \textsl{NuSTAR}/FPMA, \textsl{NuSTAR}/FPMB, and \textsl{XMM-Newton}/EPIC-Pn. The three dashed horizontal lines separate the four flux states (low, medium, high, very-high) of our spectral analysis. \label{f-mcglc}}
\end{figure}

\subsection{Data reduction}

\textsl{NuSTAR} and \textsl{XMM-Newton} observed MCG--06--30--15 simultaneously starting from 2013 January 29 for a total duration of about 360~ks. Tab.~\ref{t-obs} shows the details of the observations used in the present work. The high variability of the source is evident in its light curve in Fig.~\ref{f-mcglc}. The flux changes by a factor of 5 during these observations.

\textsl{NuSTAR}~\citep{2013ApJ...770..103H} observed MCG--06--30--15 with its focal plane modules (FPM) A and B for three consecutive observations for about 360~ks. The raw data are processed to the clean event files using the NUPIPELINE routine of the NUSTARDAS package distributed as a part of high energy analysis package HEASOFT and CALDB~v20180312. The source region is selected around the center of the source with the radius of 70~arcsec. A background region with the radius of 100~arcsec is taken on the same detector as far as possible from the source. Spectra, response files, and ancillary files are generated using the routine NUPRODUCTS. Requiring to oversample the spectral resolution by at least a factor of 2.5, we find that we need to rebin the data to have at least 70 counts per bin. More details can be found in \citet{2019ApJ...875...56T}.

\textsl{XMM-Newton}~\citep{2001A&A...365L...1J} observed MCG--06--30--15 with its EPIC-Pn and EPIC-MOS1/2 cameras for three consecutive revolutions for about 315~ks. These cameras operated in medium filter and small window mode~\citep{2001A&A...365L...1J}. We do not use the MOS data for our analysis because they are strongly affected by pile-up. The raw data for each revolution is converted into cleaned event files using SAS~v16.0.0 and then combined into single event files. Good time intervals (GTIs) are generated using TABTIGEN and then used in filtering the combined cleaned event file. For the source, a circular region with the radius of 40~arcsec is taken around its center. A background region with the radius of 50~arcsec is taken as far as possible from the source center and on the same detector. Response files and Ancillary files are generated after backscaling. The spectra are binned such that we oversample the instrument resolution by a factor of 3 and also to a minimum of 50 counts per bin in order to apply $\chi^2$-statistics. More details are presented in \citet{2019ApJ...875...56T}.

Because of the high variability of the source, strictly simultaneous flux-resolved data are used for the spectral analysis. FPMA, FPMB, and EPIC-Pn data are divided into four flux states (low, medium, high, very-high) in such a way that the photon count in every state is similar. The same flux-resolved technique was used in \citet{2019ApJ...875...56T}. \citet{2014ApJ...787...83M} adopted a different method. They grouped the data into different time intervals according to their hardness. In order to have strictly simultaneous data, the GTIs from the instruments were combined using the ftool MGTIME.

\begin{figure}
\begin{center}
\includegraphics[width=8.5cm,trim={1.5cm 0.0cm 4.0cm 17.5cm},clip]{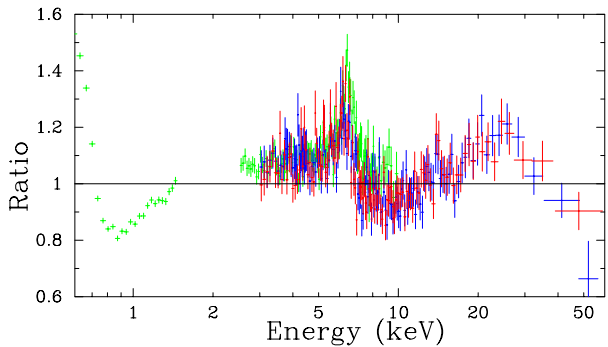}
\end{center}
\vspace{-0.7cm}
\caption{MCG--06--30--15. Data to best-fit model ratio for an absorbed power-law for the low flux state. Blue crosses are for \textsl{NuSTAR}/FPMA, red crosses for \textsl{NuSTAR}/FPMB, and green crosses for \textsl{XMM-Newton}.  \label{f-mcgr}}
\end{figure}

\subsection{Spectral analysis}

For the spectral analysis, we use XSPEC v12.11.1 \citep{1996ASPC..101...17A} distributed as a part of HEASOFT v6.28 package. We use WILMS abundances~\citep{2000ApJ...542..914W} and VERN cross-sections \citep{1996ApJ...465..487V}.

From every instrument (FPMA, FPMB, EPIC-Pn), we have four spectra corresponding to the four different flux states (low, medium, high, very-high), so twelve spectra in total. We fit all the flux states simultaneously. For each flux state, we freeze the calibration constant of EPIC-Pn to 1 and the calibration constants of FPMA ($C_{\rm FPMA}$) and FPMB ($C_{\rm FPMB}$) are left free. We find that the values of the ratio $C_{\rm FPMB}/C_{\rm FPMA}$ are always in the range 0.95 to 1.05, which agrees with the standard cross-calibration value between the two instruments.

Fig.~\ref{f-mcgr} shows the data to the best-fit model ratio when we fit the low flux state data with an absorbed power-law. Below 3~keV, we see residuals due to warm ionized absorbers~\citep{2000MNRAS.318..857L,2003ApJ...596..114S} and an excess of photon count, which is quite common in AGN \citep{2004MNRAS.349L...7G,2006MNRAS.365.1067C,2009MNRAS.394..443M,2014xru..confE.203W}. Above 3~keV, we see a prominent and broad iron line peaked around 6~keV and a Compton hump above 10~keV.

In order to fit the whole spectrum of the source, we add a relativistic reflection component, produced from the inner region of the accretion disk, and a non-relativistic reflection component, produced by illumination of some cold material far from the compact object. These components are modified by two warm ionized absorbers and a dust neutral absorber. The latter is constructed by studying the \textsl{Chandra} data of this source. A narrow emission line and a narrow absorption line are also seen in the spectra. In XSPEC language, the model reads

\vspace{0.2cm}

\noindent {\tt tbabs$\times$dustyabs$\times$warmabs$_1$$\times$warmabs$_2$$\times$(cutoffpl \\
+relxill\_nk+xillver+zgauss+zgauss)} .

\vspace{0.2cm}
 
\noindent {\tt tbabs} describes the Galactic absorption and has the hydrogen column density $N_{\rm H}$ as its only parameter~\citep{2000ApJ...542..914W}. Following \citet{1990ARA&A..28..215D}, we freeze $N_{\rm H}$ to the value $3.9\times 10^{20}$~cm$^{-2}$. {\tt dustyabs} describes a dust neutral absorber \citep[see][for the details]{2000MNRAS.318..857L}. {\tt warmabs$_1$} and {\tt warmabs$_2$} are two warm absorbers and they have a column density and an ionization as their parameters. The presence of two warm absorbers was determined in \citet{2019ApJ...875...56T} comparing the $\chi^2$ of models with a different number of warm absorbers. We also note that \citet{2014ApJ...787...83M} found two warm absorbers in this data set from the analysis of RGS data.

{\tt cutoffpl} models the primary emission from the Comptonized corona with an exponential high-energy cutoff. {\tt relxill\_nk} describes the relativistic reflection component~\citep{2017ApJ...842...76B,2019ApJ...878...91A}. In this paper, we use the version in which the spacetime is described by the Johannsen metric with non-vanishing deformation parameter $\alpha_{13}$~\citep{2013PhRvD..88d4002J} and the accretion disk has a finite thickness~\citep{2020ApJ...899...80A}. {\tt xillver} describes the non-relativistic reflection component generated by illumination by the corona of some cold material relatively far from the central black hole~\citep{2010ApJ...718..695G}.

{\tt zgauss} is a redshifted Gaussian profile and is used to fit two remaining prominent features. One of them can be interpreted as a narrow emission oxygen line at 0.81~keV. The other feature is a narrow absorption line at 1.22~keV and can be interpreted as a blueshifted oxygen absorption line due to some relativistic outflow~\citep{1997ApJ...489L..25L}. These two lines are relatively prominent in every flux state; see the single flux state analysis in \citet{2019ApJ...875...56T}, in particular their Tab.~5. We do not see any significant energy shift among different flux states. The normalization of the 0.81~keV line is of the order of $10^{-2}$, while the 1.22~keV line is less prominent and its normalization is of the order of $10^{-5}$. These features were found even in \citet{2014ApJ...787...83M}, see their Fig.~7, as well as in the analysis of the RGS data, see their Fig.~6. As they are narrow features at low energy, their inclusion/exclusion has no impact on the estimates of the black hole spin and of the deformation parameter; Including the two Gaussians we just decrease the $\chi^2$.

We fit the four flux states together. For \textsl{NuSTAR}, we use the FPMA and FPMB data in the energy range 3-80~keV. For \textsl{XMM-Newton}, we use the data in the energy range 0.5-10.0~keV, but we exclude the data in the energy range 1.5-2.5~keV because of instrumental issues as discussed in~\citet{2014ApJ...787...83M} and \citet{2019ApJ...875...56T}. The data below 0.5~keV and above 10.0~keV are ignored because of their poor quality and strong background, respectively.

\begin{table*}
\centering
{\renewcommand{\arraystretch}{1.3}
\begin{tabular}{l|cccc|cccc}
\hline\hline
 & \multicolumn{4}{c}{Model~0} & \multicolumn{4}{c}{Model~1} \\
\hline\hline
Group & 1 & 2 & 3 & 4 & 1 & 2 & 3 & 4 \\
\hline
{\tt tbabs} &&&& &&&& \\
$N_{\rm H} / 10^{22}$~cm$^{-2}$ & \multicolumn{4}{c}{$0.039^\star$} & \multicolumn{4}{c}{$0.039^\star$} \\
\hline
{\tt warmabs$_1$} &&&& \\
$N_{\rm H \, 1} / 10^{22}$ cm$^{-2}$ & $0.49^{+0.22}_{-0.07}$ & $1.180^{+0.018}_{-0.025}$ & $1.007^{+0.020}_{-0.027}$ & $0.74^{+0.10}_{-0.04}$ 
& $0.65^{+0.05}_{-0.11}$ & $1.19^{+0.03}_{-0.05}$ & $1.02^{+0.04}_{-0.05}$ & $0.75^{+0.04}_{-0.07}$ \\
$\log\xi_1$ & $1.86^{+0.04}_{-0.04}$ & $1.954^{+0.013}_{-0.020}$ & $1.920^{+0.015}_{-0.019}$ & $1.830^{+0.013}_{-0.019}$ 
& $1.91^{+0.04}_{-0.08}$ & $1.953^{+0.016}_{-0.018}$ & $1.917^{+0.020}_{-0.021}$ & $1.83^{+0.04}_{-0.03}$ \\
\hline
{\tt warmabs$_2$} &&&& \\
$N_{\rm H \, 2} / 10^{22}$ cm$^{-2}$ & $0.63^{+2.06}_{-0.06}$ & $0.02^{+0.02}_{-0.02}$ & $0.52^{+0.18}_{-0.14}$ & $0.25^{+0.05}_{-0.11}$ 
& $0.47^{+0.06}_{-0.05}$ & $0.02^{+0.02}_{-0.02}$ & $0.51^{+0.18}_{-0.18}$ & $0.25^{+0.06}_{-0.05}$ \\
$\log\xi_2$ & $1.90^{+0.03}_{-0.08}$ & $3.1_{-0.6}$ & $3.23^{+0.06}_{-0.07}$ & $2.48^{+0.13}_{-0.13}$ 
& $1.85^{+0.08}_{-0.04}$ & $3.1_{-1.0}$ & $3.22^{+0.07}_{-0.11}$ & $2.48^{+0.10}_{-0.14}$ \\
\hline
{\tt dustyabs} &&&& \\
$\log \big( N_{\rm Fe} / 10^{21}$ cm$^{-2} \big)$ & \multicolumn{4}{c}{$17.408^{+0.018}_{-0.031}$} & \multicolumn{4}{c}{$17.399^{+0.031}_{-0.014}$} \\
\hline
{\tt cutoffpl} &&&& \\
$\Gamma$ & $1.954^{+0.012}_{-0.014}$ & $1.973^{+0.008}_{-0.006}$ & $2.015^{+0.006}_{-0.007}$ & $2.027^{+0.006}_{-0.007}$ & $1.953^{+0.007}_{-0.008}$ & $1.973^{+0.010}_{-0.011}$ & $2.018^{+0.013}_{-0.012}$ & $2.030^{+0.011}_{-0.012}$ \\
$E_{\rm cut}$ [keV] & $196^{+36}_{-28}$ & $154^{+33}_{-21}$ & $164^{+33}_{-23}$ & $279^{+137}_{-59}$
& $200^{+56}_{-41}$ & $156^{+45}_{-28}$ & $167^{+48}_{-33}$ & $290^{+83}_{-86}$ \\
norm~$(10^{-3})$ & $8.4^{+0.4}_{-0.4}$ & $12.32^{+0.22}_{-0.78}$ & $15.3^{+1.2}_{-0.7}$ & $21.1^{+0.8}_{-1.0}$ & $8.3^{+0.3}_{-0.3}$ & $12.3^{+0.4}_{-0.7}$ & $15.7^{+0.5}_{-0.6}$ & $21.6^{+0.7}_{-1.1}$ \\ 
\hline
{\tt relxill\_nk} &&&& \\
$\dot{m}$ & \multicolumn{4}{c}{$0^\star$} & \multicolumn{4}{c}{$0.05^\star$} \\
$q_{\rm in}$ & $6.2^{+1.6}_{-1.5}$ & $7.1^{+1.4}_{-1.5}$ & $7.9^{+0.7}_{-0.5}$ & $8.7^{+0.7}_{-0.5}$
& $7.7^{+2.2}_{-4.1}$ & $8.1^{+1.8}_{-3.6}$ & $8.7^{+1.0}_{-1.2}$ & $10.0_{-1.1}$ \\
$q_{\rm out}$ & \multicolumn{4}{c}{$3^\star$} & \multicolumn{4}{c}{$3^\star$} \\
$R_{\rm br}$ [$M$] & $2.9^{+0.4}_{-0.4}$ & $3.0^{+0.3}_{-0.3}$ & $3.24^{+0.16}_{-0.11}$ & $3.27^{+0.14}_{-0.17}$
& $2.8^{+0.4}_{-0.4}$ & $2.83^{+0.68}_{-0.19}$ & $3.15^{+0.39}_{-0.18}$ & $3.1^{+0.3}_{-0.3}$ \\
$\iota$ [deg] & \multicolumn{4}{c}{$31.4^{+1.3}_{-1.4}$} & \multicolumn{4}{c}{$31.4^{+1.5}_{-1.5}$} \\
$a_*$ & \multicolumn{4}{c}{$0.972^{+0.012}_{-0.029}$} & \multicolumn{4}{c}{$0.95^{\rm + (P)}_{-0.04}$} \\
$\alpha_{13}$ & \multicolumn{4}{c}{$0.10_{-0.30}^{+0.05}$} & \multicolumn{4}{c}{$0.0^{+0.3}_{-0.4}$} \\
$z$ & \multicolumn{4}{c}{$0.007749^\star$} & \multicolumn{4}{c}{$0.007749^\star$} \\
$\log\xi$ & $2.87^{+0.05}_{-0.06}$ & $3.004^{+0.008}_{-0.078}$ & $3.058^{+0.011}_{-0.008}$ & $3.136^{+0.024}_{-0.027}$ 
& $2.89^{+0.07}_{-0.06}$ & $3.01^{+0.03}_{-0.11}$ & $3.052^{+0.023}_{-0.021}$ & $3.13^{+0.04}_{-0.04}$ \\
$A_{\rm Fe}$ & \multicolumn{4}{c}{$3.1^{+0.3}_{-0.3}$} & \multicolumn{4}{c}{$3.0^{+0.3}_{-0.3}$} \\
norm~$(10^{-3})$ & $0.049^{+0.003}_{-0.004}$ & $0.061^{+0.006}_{-0.004}$ & $0.103^{+0.005}_{-0.006}$ & $0.130^{+0.006}_{-0.009}$ & $0.048^{+0.005}_{-0.003}$ & $0.060^{+0.004}_{-0.005}$ & $0.097^{+0.008}_{-0.008}$ & $0.122^{+0.014}_{-0.007}$ \\ 
\hline
{\tt xillver} &&&& \\
$\log\xi'$ & \multicolumn{4}{c}{$0^\star$} & \multicolumn{4}{c}{$0^\star$} \\
norm~$(10^{-3})$ & \multicolumn{4}{c}{$0.058^{+0.005}_{-0.005}$} & \multicolumn{4}{c}{$0.058^{+0.007}_{-0.008}$} \\
\hline
{\tt zgauss} &&&& &&&&\\
$E_{\rm line}$ [keV] & \multicolumn{4}{c}{$0.8143^{+0.0004}_{-0.0024}$} & \multicolumn{4}{c}{$0.8141^{+0.0001}_{-0.0045}$} \\
\hline
{\tt zgauss} &&&& &&&&\\
$E_{\rm line}$ [keV] & \multicolumn{4}{c}{$1.265^{+0.011}_{-0.009}$} & \multicolumn{4}{c}{$1.225^{+0.012}_{-0.108}$} \\
\hline
$\chi^2$/dof & \multicolumn{4}{c}{$3027.96/2685 = 1.12773$} & \multicolumn{4}{c}{$3028.67/2685 = 1.12800$} \\
\hline\hline
\end{tabular}
}
\vspace{0.2cm}
\caption{\rm MCG--06--30--15. Summary of the best-fit values for Model~0 ($\dot{m} = 0$) and Model~1 ($\dot{m} = 0.05$). The ionization parameters ($\xi$, $\xi_1$, $\xi_2$, and $\xi'$) are in units erg~cm~s$^{-1}$. The reported uncertainties correspond to the 90\% confidence level for one relevant parameter ($\Delta\chi^2 = 2.71$). $^\star$ indicates that the value of the parameter is frozen in the fit. (P) means that the 90\% confidence level reaches the upper boundary of the black hole spin parameter ($a_*^{\rm max} = 0.998$). $q_{\rm in}$ is allowed to vary in the range 0 to 10. \label{t-mcg-1}}
\end{table*}

\begin{table*}
\centering
{\renewcommand{\arraystretch}{1.3}
\begin{tabular}{l|cccc|cccc}
\hline\hline
 & \multicolumn{4}{c}{Model~2} & \multicolumn{4}{c}{Model~3} \\
\hline\hline
Group & 1 & 2 & 3 & 4 & 1 & 2 & 3 & 4 \\
\hline
{\tt tbabs} &&&& &&&& \\
$N_{\rm H} / 10^{22}$~cm$^{-2}$ & \multicolumn{4}{c}{$0.039^\star$} & \multicolumn{4}{c}{$0.039^\star$} \\
\hline
{\tt warmabs$_1$} &&&& \\
$N_{\rm H \, 1} / 10^{22}$ cm$^{-2}$ & $0.49^{+0.06}_{-0.98}$ & $1.174^{+0.011}_{-0.044}$ & $1.00^{+0.04}_{-0.03}$ & $0.74^{+0.09}_{-0.02}$ 
& $0.63^{+0.12}_{-0.76}$ & $1.182^{+0.021}_{-0.062}$ & $1.013^{+0.024}_{-0.044}$ & $0.74^{+0.07}_{-0.05}$ \\
$\log\xi_1$ & $1.86^{+0.03}_{-0.04}$ & $1.956^{+0.006}_{-0.020}$ & $1.921^{+0.006}_{-0.023}$ & $1.830^{+0.008}_{-0.018}$ 
& $1.90^{+0.03}_{-0.07}$ & $1.956^{+0.015}_{-0.026}$ & $1.920^{+0.016}_{-0.022}$ & $1.829^{+0.022}_{-0.021}$ \\
\hline
{\tt warmabs$_2$} &&&& \\
$N_{\rm H \, 2} / 10^{22}$ cm$^{-2}$ & $0.62^{+2.09}_{-0.04}$ & $0.02^{+0.02}_{-0.02}$ & $0.53^{+0.14}_{-0.07}$ & $0.25^{+0.03}_{-0.05}$ 
& $0.49^{+3.64}_{-0.06}$ & $0.02^{+0.02}_{-0.02}$ & $0.52^{+0.18}_{-0.17}$ & $0.25^{+0.05}_{-0.05}$ \\
$\log\xi_2$ & $1.906^{+0.012}_{-0.084}$ & $3.1_{-0.5}$ & $3.23^{+0.04}_{-0.04}$ & $2.48^{+0.09}_{-0.14}$ 
& $1.860^{+0.05}_{-0.04}$ & $3.1_{-0.7}$ & $3.23^{+0.06}_{-0.08}$ & $2.48^{+0.13}_{-0.13}$ \\
\hline
{\tt dustyabs} &&&& \\
$\log \big( N_{\rm Fe} / 10^{21}$ cm$^{-2} \big)$ & \multicolumn{4}{c}{$17.410^{+0.018}_{-0.032}$} & \multicolumn{4}{c}{$17.405^{+0.017}_{-0.028}$} \\
\hline
{\tt cutoffpl} &&&& \\
$\Gamma$ & $1.953^{+0.006}_{-0.013}$ & $1.971^{+0.004}_{-0.010}$ & $2.014^{+0.003}_{-0.011}$ & $2.027^{+0.003}_{-0.011}$ & $1.953^{+0.011}_{-0.007}$ & $1.971^{+0.011}_{-0.014}$ & $2.015^{+0.011}_{-0.009}$ & $2.028^{+0.009}_{-0.009}$ \\
$E_{\rm cut}$ [keV] & $196^{+9}_{-24}$ & $155^{+9}_{-13}$ & $163^{+10}_{-10}$ & $269^{+135}_{-72}$
& $197^{+44}_{-37}$ & $155^{+38}_{-23}$ & $164^{+38}_{-26}$ & $280^{+159}_{-69}$ \\
norm~$(10^{-3})$ & $8.43^{+0.07}_{-0.03}$ & $12.12^{+0.07}_{-0.35}$ & $15.3^{+1.2}_{-0.6}$ & $21.2^{+1.7}_{-1.2}$ & $8.4^{+0.5}_{-0.5}$ & $12.1^{+0.8}_{-0.5}$ & $15.5^{+1.2}_{-1.2}$ & $21.4^{+0.7}_{-1.1}$ \\ 
\hline
{\tt relxill\_nk} &&&& \\
$\dot{m}$ & \multicolumn{4}{c}{$0.1^\star$} & \multicolumn{4}{c}{$0.2^\star$} \\
$q_{\rm in}$ & $6.4^{+0.3}_{-1.1}$ & $7.70^{+0.21}_{-0.22}$ & $7.6^{+1.0}_{-1.3}$ & $8.9^{+0.7}_{-0.9}$
& $7.0^{+1.7}_{-2.5}$ & $8.0^{+0.4}_{-0.3}$ & $8.2^{+0.8}_{-0.7}$ & $8.36^{+0.63}_{-0.15}$ \\
$q_{\rm out}$ & \multicolumn{4}{c}{$3^\star$} & \multicolumn{4}{c}{$3^\star$} \\
$R_{\rm br}$ [$M$] & $2.95^{+0.07}_{-0.26}$ & $2.94^{+0.48}_{-0.04}$ & $3.33^{+0.03}_{-0.16}$ & $3.36^{+0.15}_{-0.02}$
& $3.0^{+0.3}_{-0.5}$ & $3.0^{+0.4}_{-0.3}$ & $3.3^{+0.3}_{-0.3}$ & $3.35^{+0.33}_{-0.14}$ \\
$\iota$ [deg] & \multicolumn{4}{c}{$31.5^{+1.1}_{-1.5}$} & \multicolumn{4}{c}{$31.5^{+1.4}_{-1.6}$} \\
$a_*$ & \multicolumn{4}{c}{$0.964^{+0.016}_{-0.025}$} & \multicolumn{4}{c}{$0.963^{\rm + (P)}_{-0.028}$} \\
$\alpha_{13}$ & \multicolumn{4}{c}{$-0.01^{+0.11}_{-0.27}$} & \multicolumn{4}{c}{$0.0^{+0.25}_{-0.25}$} \\
$z$ & \multicolumn{4}{c}{$0.007749^\star$} & \multicolumn{4}{c}{$0.007749^\star$} \\
$\log\xi$ & $2.87^{+0.03}_{-0.07}$ & $3.009^{+0.002}_{-0.040}$ & $3.057^{+0.021}_{-0.012}$ & $3.133^{+0.033}_{-0.022}$ 
& $2.88^{+0.05}_{-0.09}$ & $3.009^{+0.013}_{-0.068}$ & $3.056^{+0.016}_{-0.019}$ & $3.13^{+0.03}_{-0.03}$ \\
$A_{\rm Fe}$ & \multicolumn{4}{c}{$3.11^{+0.24}_{-0.13}$} & \multicolumn{4}{c}{$3.08^{+0.25}_{-0.22}$} \\
norm~$(10^{-3})$ & $0.049^{+0.002}_{-0.004}$ & $0.063^{+0.001}_{-0.006}$ & $0.102^{+0.009}_{-0.001}$ & $0.126^{+0.019}_{-0.010}$ & $0.049^{+0.003}_{-0.003}$ & $0.062^{+0.004}_{-0.006}$ & $0.100^{+0.005}_{-0.005}$ & $0.125^{+0.013}_{-0.017}$ \\ 
\hline
{\tt xillver} &&&& \\
$\log\xi'$ & \multicolumn{4}{c}{$0^\star$} & \multicolumn{4}{c}{$0^\star$} \\
norm~$(10^{-3})$ & \multicolumn{4}{c}{$0.057^{+0.003}_{-0.004}$} & \multicolumn{4}{c}{$0.057^{+0.006}_{-0.006}$} \\
\hline
{\tt zgauss} &&&& &&&&\\
$E_{\rm line}$ [keV] & \multicolumn{4}{c}{$0.8143^{+0.0007}_{-0.0012}$} & \multicolumn{4}{c}{$0.8142^{+0.0001}_{-0.0003}$} \\
\hline
{\tt zgauss} &&&& &&&&\\
$E_{\rm line}$ [keV] & \multicolumn{4}{c}{$1.227^{+0.013}_{-0.010}$} & \multicolumn{4}{c}{$1.226^{+0.012}_{-0.010}$} \\
\hline
$\chi^2$/dof & \multicolumn{4}{c}{$3027.55/2685 = 1.12758$} & \multicolumn{4}{c}{$3027.47/2685 = 1.12755$} \\
\hline\hline
\end{tabular}
}
\vspace{0.2cm}
\caption{\rm MCG--06--30--15. As in Tab.~\ref{t-mcg-1} for Model~2 ($\dot{m} = 0.1$) and Model~3 ($\dot{m} = 0.2$). \label{t-mcg-2}}
\end{table*}

\begin{table*}
\centering
{\renewcommand{\arraystretch}{1.3}
\begin{tabular}{l|cccc}
\hline\hline
 & \multicolumn{4}{c}{Model~4} \\
\hline\hline
Group & 1 & 2 & 3 & 4 \\
\hline
{\tt tbabs} &&&& \\
$N_{\rm H} / 10^{22}$~cm$^{-2}$ & \multicolumn{4}{c}{$0.039^\star$} \\
\hline
{\tt warmabs$_1$} &&&& \\
$N_{\rm H \, 1} / 10^{22}$ cm$^{-2}$ & $0.52^{+2.72}_{-0.06}$ & $1.178^{+0.024}_{-0.032}$ & $1.01^{+0.03}_{-0.03}$ & $0.74^{+0.12}_{-0.05}$ \\
$\log\xi_1$ & $1.87^{+0.05}_{-0.03}$ & $1.955^{+0.015}_{-0.020}$ & $1.921^{+0.019}_{-0.023}$ & $1.83^{+0.03}_{-0.03}$ \\
\hline
{\tt warmabs$_2$} &&&& \\
$N_{\rm H \, 2} / 10^{22}$ cm$^{-2}$ & $0.59^{+2.97}_{-0.07}$ & $0.02^{+0.02}_{-0.02}$ & $0.53^{+0.19}_{-0.18}$ & $0.26^{+0.06}_{-0.05}$ \\
$\log\xi_2$ & $1.90^{+0.04}_{-0.07}$ & $3.1_{-0.8}$ & $3.23^{+0.07}_{-0.09}$ & $2.48^{+0.11}_{-0.15}$ \\
\hline
{\tt dustyabs} &&&& \\
$\log \big( N_{\rm Fe} / 10^{21}$ cm$^{-2} \big)$ & \multicolumn{4}{c}{$17.411^{+0.010}_{-0.031}$} \\
\hline
{\tt cutoffpl} &&&& \\
$\Gamma$ & $1.954^{+0.011}_{-0.008}$ & $1.974^{+0.013}_{-0.010}$ & $2.014^{+0.013}_{-0.010}$ & $2.028^{+0.012}_{-0.010}$ \\
$E_{\rm cut}$ [keV] & $200^{+44}_{-38}$ & $149^{+39}_{-22}$ & $161^{+38}_{-28}$ & $274^{+148}_{-73}$\\
norm~$(10^{-3})$ & $8.48^{+0.15}_{-0.26}$ & $12.5^{+0.4}_{-0.5}$ & $15.5^{+0.5}_{-0.6}$ & $21.5^{+0.7}_{-1.7}$\\ 
\hline
{\tt relxill\_nk} &&&& \\
$\dot{m}$ & \multicolumn{4}{c}{$0.3^\star$} \\
$q_{\rm in}$ & $7.5^{+1.9}_{-4.7}$ & $7.9^{+1.8}_{-2.4}$ & $8.6^{+0.7}_{-0.8}$ & $9.1^{+0.7}_{-0.7}$\\
$q_{\rm out}$ & \multicolumn{4}{c}{$3^\star$}\\
$R_{\rm br}$ [$M$] & $2.9^{+0.3}_{-0.5}$ & $2.9^{+0.3}_{-0.4}$ & $3.19^{+0.17}_{-0.18}$ & $3.28^{+0.19}_{-0.16}$\\
$\iota$ [deg] & \multicolumn{4}{c}{$31.8^{+1.4}_{-1.6}$}\\
$a_*$ & \multicolumn{4}{c}{$0.953^{+0.033}_{-0.021}$}\\
$\alpha_{13}$ & \multicolumn{4}{c}{$-0.17^{+0.27}_{-0.23}$}\\
$z$ & \multicolumn{4}{c}{$0.007749^\star$}\\
$\log\xi$ & $2.87^{+0.05}_{-0.06}$ & $2.96^{+0.05}_{-0.09}$ & $3.055^{+0.020}_{-0.017}$ & $3.13^{+0.04}_{-0.03}$ \\
$A_{\rm Fe}$ & \multicolumn{4}{c}{$3.2^{+0.3}_{-0.4}$}\\
norm~$(10^{-3})$ & $0.049^{+0.004}_{-0.004}$ & $0.059^{+0.007}_{-0.004}$ & $0.101^{+0.005}_{-0.008}$ & $0.124^{+0.009}_{-0.007}$\\ 
\hline
{\tt xillver} &&&& \\
$\log\xi'$ & \multicolumn{4}{c}{$0^\star$} \\
norm~$(10^{-3})$ & \multicolumn{4}{c}{$0.058^{+0.007}_{-0.007}$}\\
\hline
{\tt zgauss} &&&&\\
$E_{\rm line}$ [keV] & \multicolumn{4}{c}{$0.8142^{+0.0007}_{-0.0009}$}\\
\hline
{\tt zgauss} &&&&\\
$E_{\rm line}$ [keV] & \multicolumn{4}{c}{$1.226^{+0.011}_{-0.009}$}\\
\hline
$\chi^2$/dof & \multicolumn{4}{c}{$3027.71/2685 = 1.12764$} \\
\hline\hline
\end{tabular}
}
\vspace{0.2cm}
\caption{\rm MCG--06--30--15. As in Tab.~\ref{t-mcg-1} for Model~4 ($\dot{m} = 0.3$). \label{t-mcg-3}}
\end{table*}

The hydrogen column density $N_{\rm H}$ in {\tt tbabs} is frozen to the same value for all flux states while the iron column density in {\tt dustyabs} is free in the fit but is still kept constant over the four flux states, as there are no reasons to have any appreciable variation of its value over the timescale of the observation. See \citet{2000MNRAS.318..857L} for more details about the absorption of this source. In the two warm absorbers, {\tt warmabs$_1$} and {\tt warmabs$_2$}, their column density and ionization are free in the fit and are allowed to vary among different flux states: these are warm ionized clouds around the source and their timescale variability can be short. Even the primary emission from the corona, here described by {\tt cutoffpl}, can change over short timescales, so its photon index, high-energy cutoff, and normalization are all free in the fit and are allowed to vary between different flux states.

In {\tt relxill\_nk}, the emissivity profile is described by a broken power-law and we have thus three parameters: inner emissivity index $q_{\rm in}$, outer emissivity index $q_{\rm out}$, and breaking radius $R_{\rm br}$. Initially they are all free in the fit and are allowed to vary between different flux states, as the illumination of the disk may change from one flux state to another flux state. However, we find that $q_{\rm out}$ is close to 3 in all flux states. We thus repeat the fit with the outer emissivity index frozen to 3 in all flux states. The normalization of {\tt relxill\_nk} is allowed to vary among different flux states, as the reflection component depends on the primary emission from the corona, which may be different among different flux states. The reflection fraction in {\tt relxill\_nk} is frozen to $-1$ as the continuum from the corona is already described by {\tt cutoffpl}. The parameters that are not thought to vary over the timescale of the observation have their value tied between different flux states: black hole spin parameter $a_*$, viewing angle $\iota$, deformation parameter $\alpha_{13}$, and iron abundance $A_{\rm Fe}$. All other free parameters in the model are allowed to vary between different flux states, including the ionization parameter $\xi$, as the illumination of the accretion disk can change too. The mass accretion rate, $\dot{m}$, which is the parameter regulating the thickness of the disk, is frozen to 0 (Model~0), 0.05 (Model~1), 0.1 (Model~2), 0.2 (Model~3), and 0.3 (Model~4).

We impose the same value of normalization of {\tt xillver} for all flux states as the emission of the distant reflector is not expected to vary substantially between different flux states. Its values of the photon index and high-energy cutoff are tied to the values of these parameters in {\tt cutoffpl} and are allowed to be different for different flux states. The ionization parameter in {\tt xillver} is frozen to 0 as the distant reflector is supposed to be far from the central black hole and the material should be neutral. The iron abundance of the distant reflector is frozen to the solar value.

\begin{figure*}
\begin{center}
\includegraphics[width=8.5cm,trim={2.3cm 0.0cm 2.5cm 18.0cm},clip]{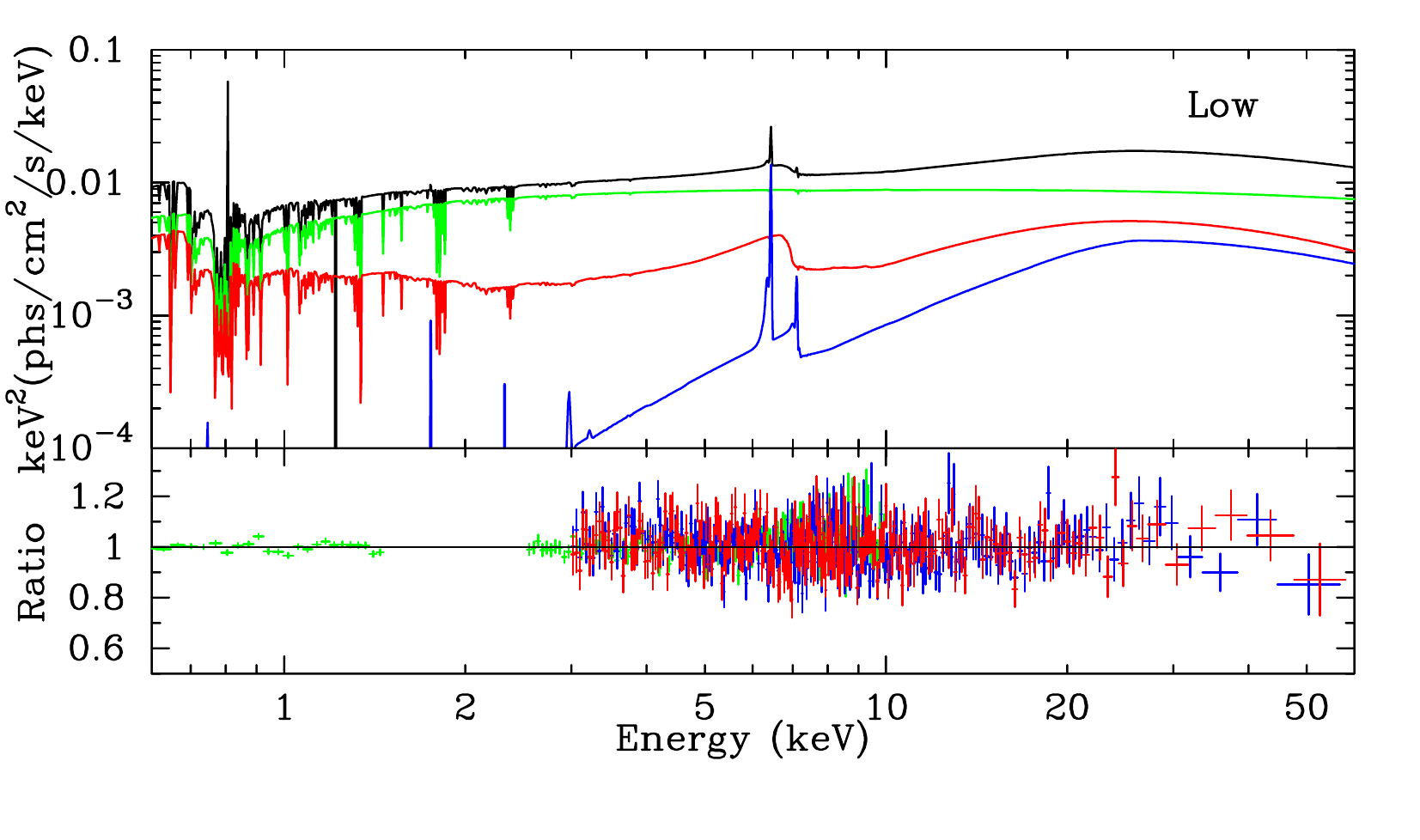}
\includegraphics[width=8.5cm,trim={2.3cm 0.0cm 2.5cm 18.0cm},clip]{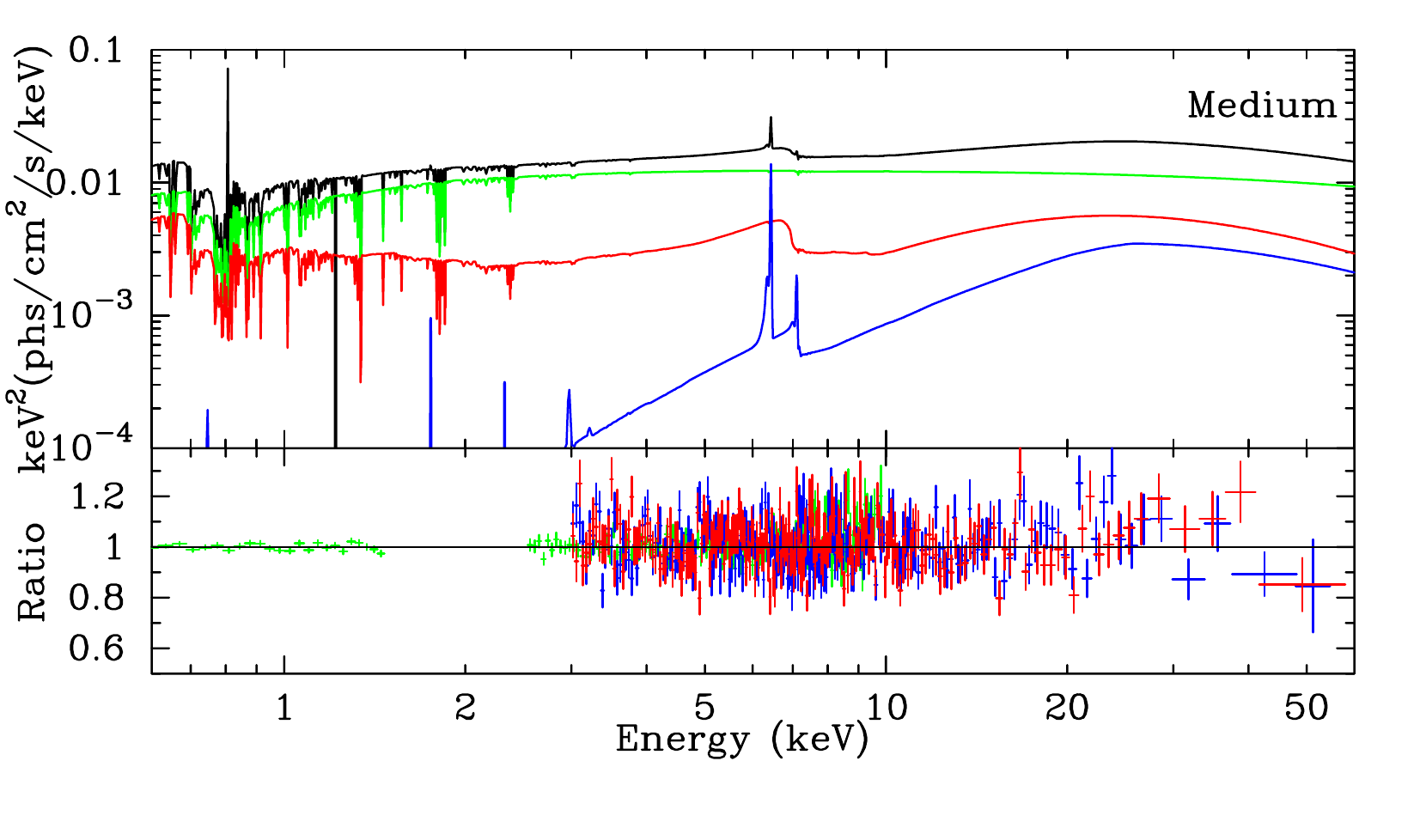} \\
\includegraphics[width=8.5cm,trim={2.3cm 0.0cm 2.5cm 18.0cm},clip]{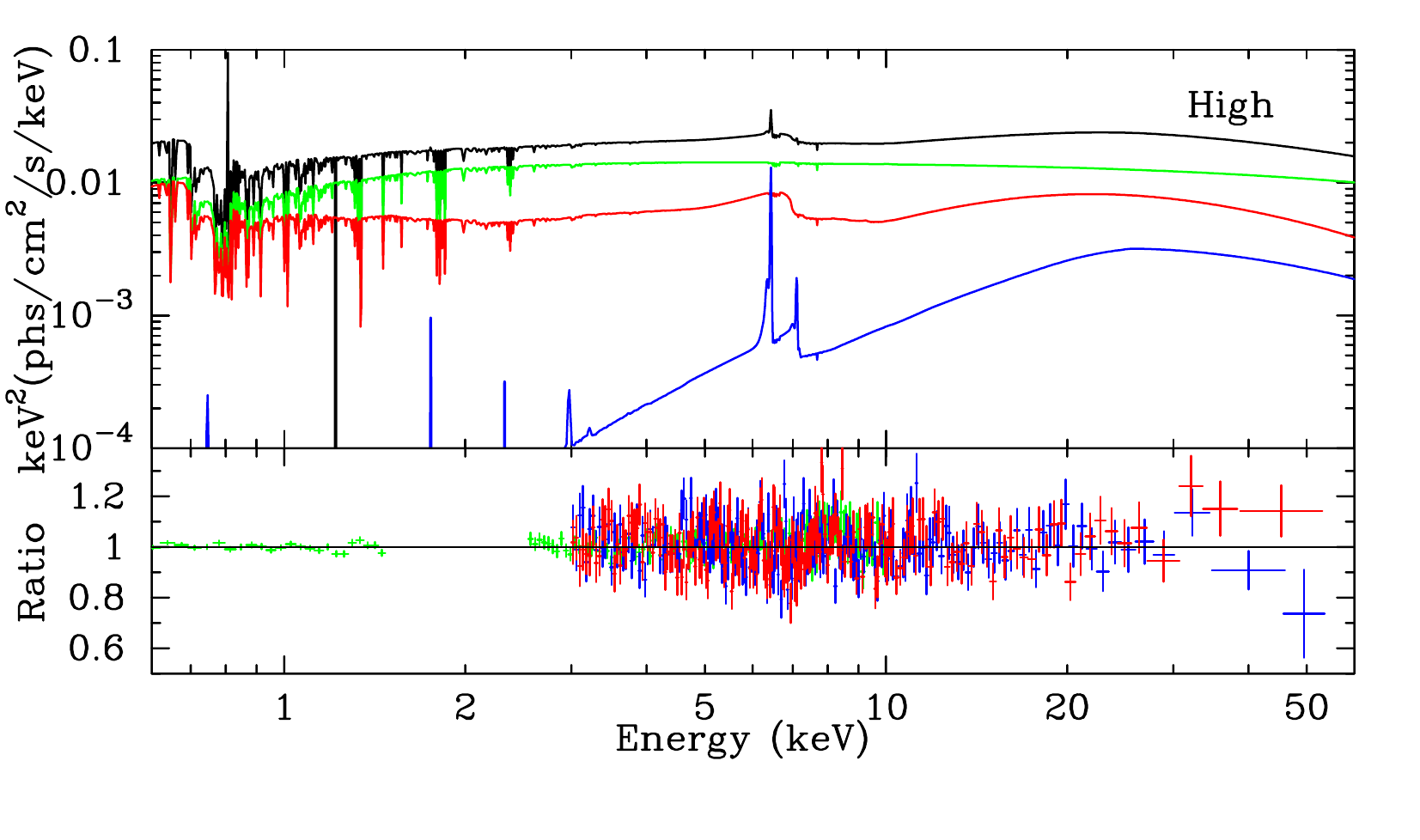}
\includegraphics[width=8.5cm,trim={2.3cm 0.0cm 2.5cm 18.0cm},clip]{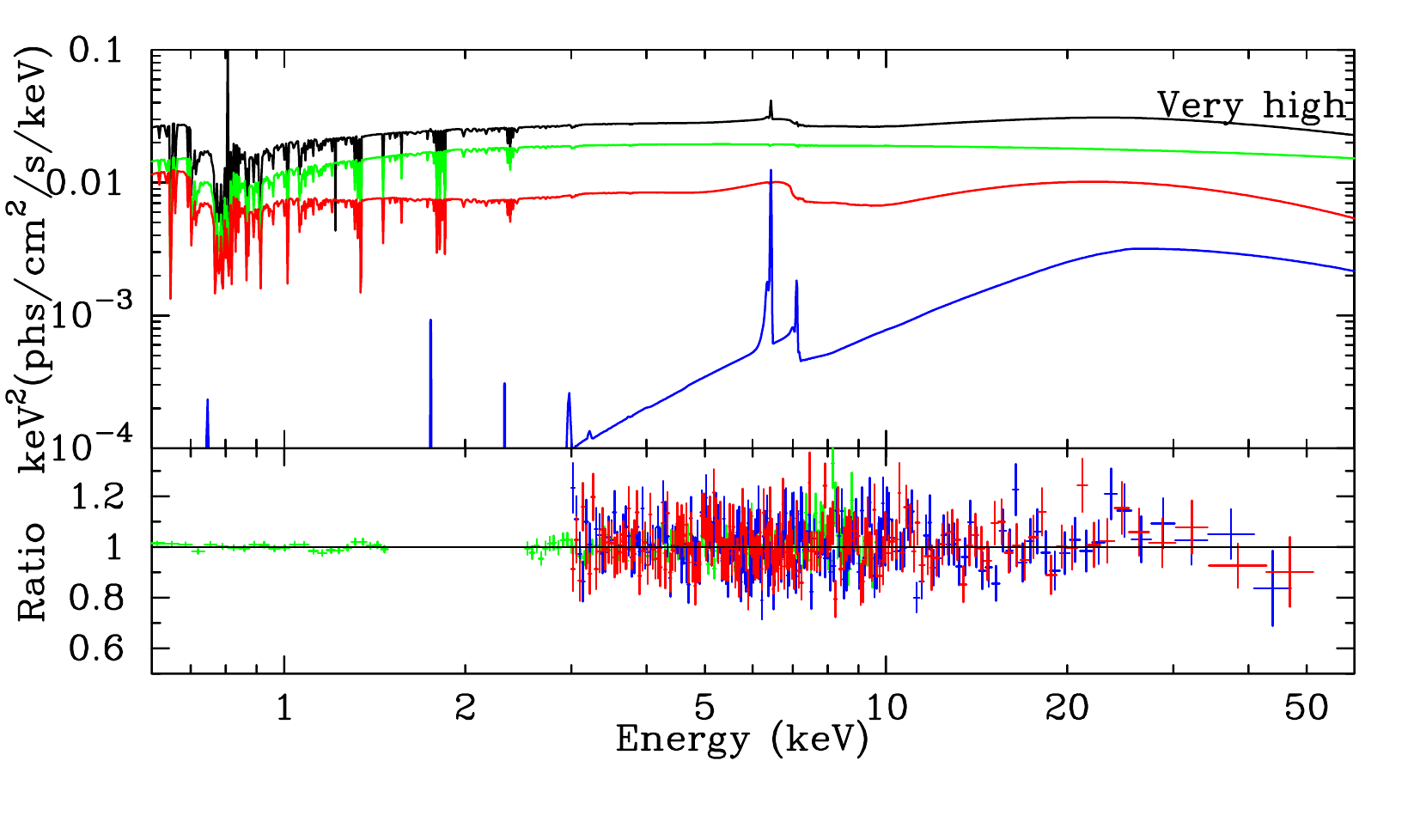}
\end{center}
\vspace{-0.7cm}
\caption{Best-fit models and data to the best-fit model ratios for the four flux states of MCG--06--30--15 for Model~0 with an infinitesimally thin accretion disk. In the best-fit model plots, the total spectrum is in black, the {\tt cutoffpl} component is in green, the {\tt relxill\_nk} component is in red, and the {\tt xillver} component is in blue. In the ratio plots, blue crosses are for \textsl{NuSTAR}/FPMA, red crosses for \textsl{NuSTAR}/FPMB, and green crosses for \textsl{XMM-Newton}.
\label{f-mcgm}}
\end{figure*}

\begin{figure*}
\begin{center}
\includegraphics[width=8.5cm,trim={0.5cm 1.5cm 0.5cm 0.5cm},clip]{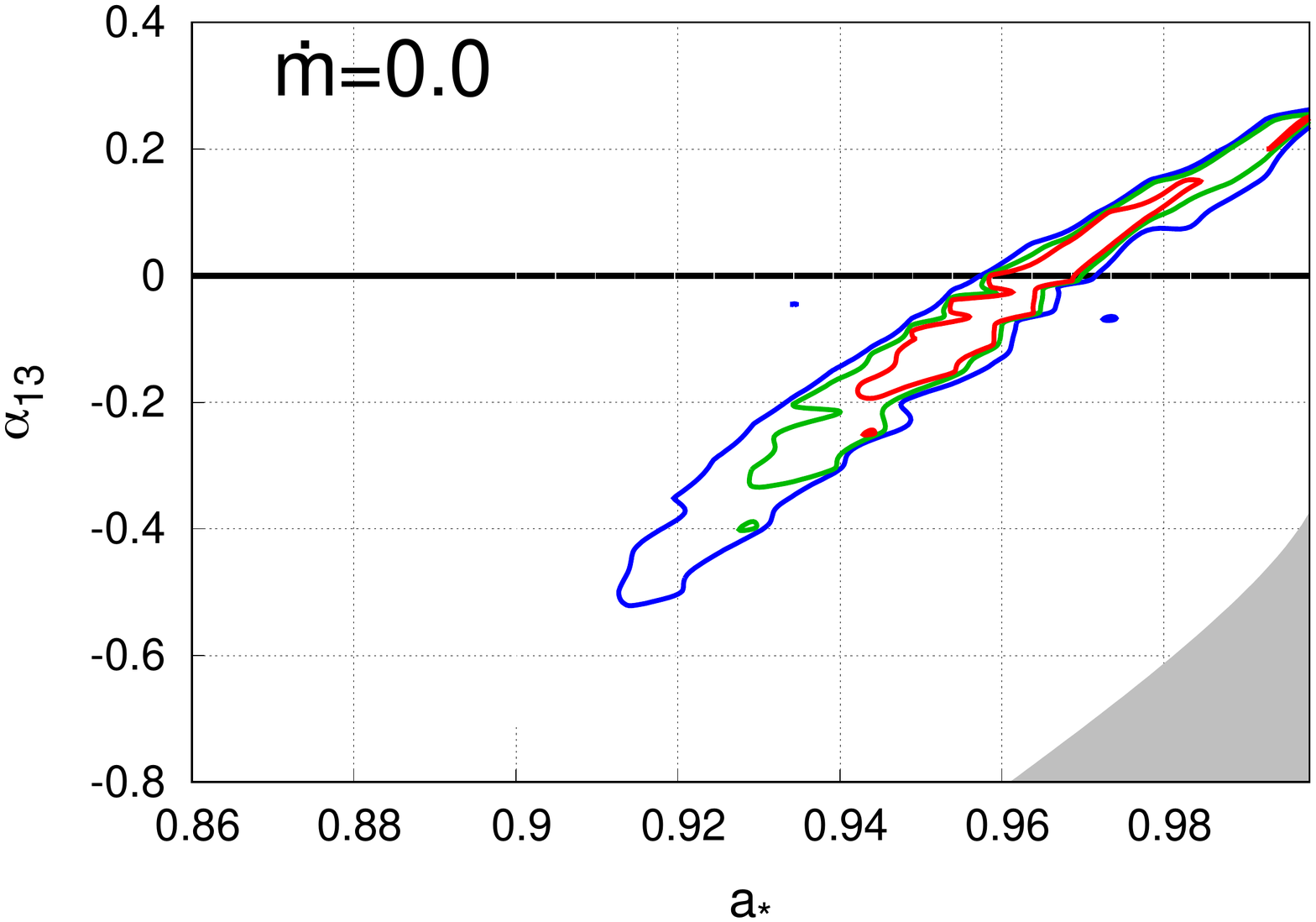} \\
\includegraphics[width=8.5cm,trim={0.5cm 1.5cm 0.5cm 0.5cm},clip]{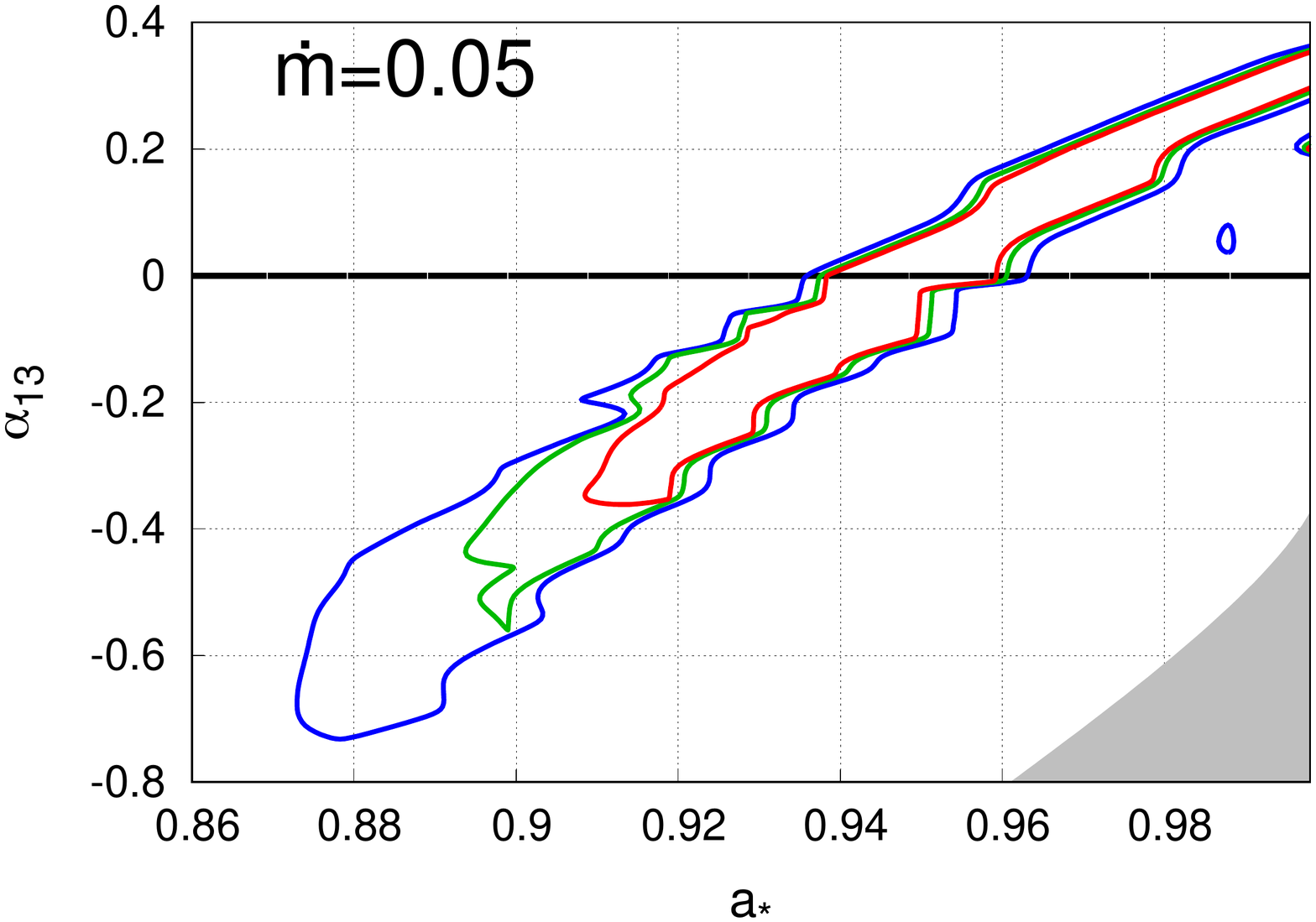}
\includegraphics[width=8.5cm,trim={0.5cm 1.5cm 0.5cm 0.5cm},clip]{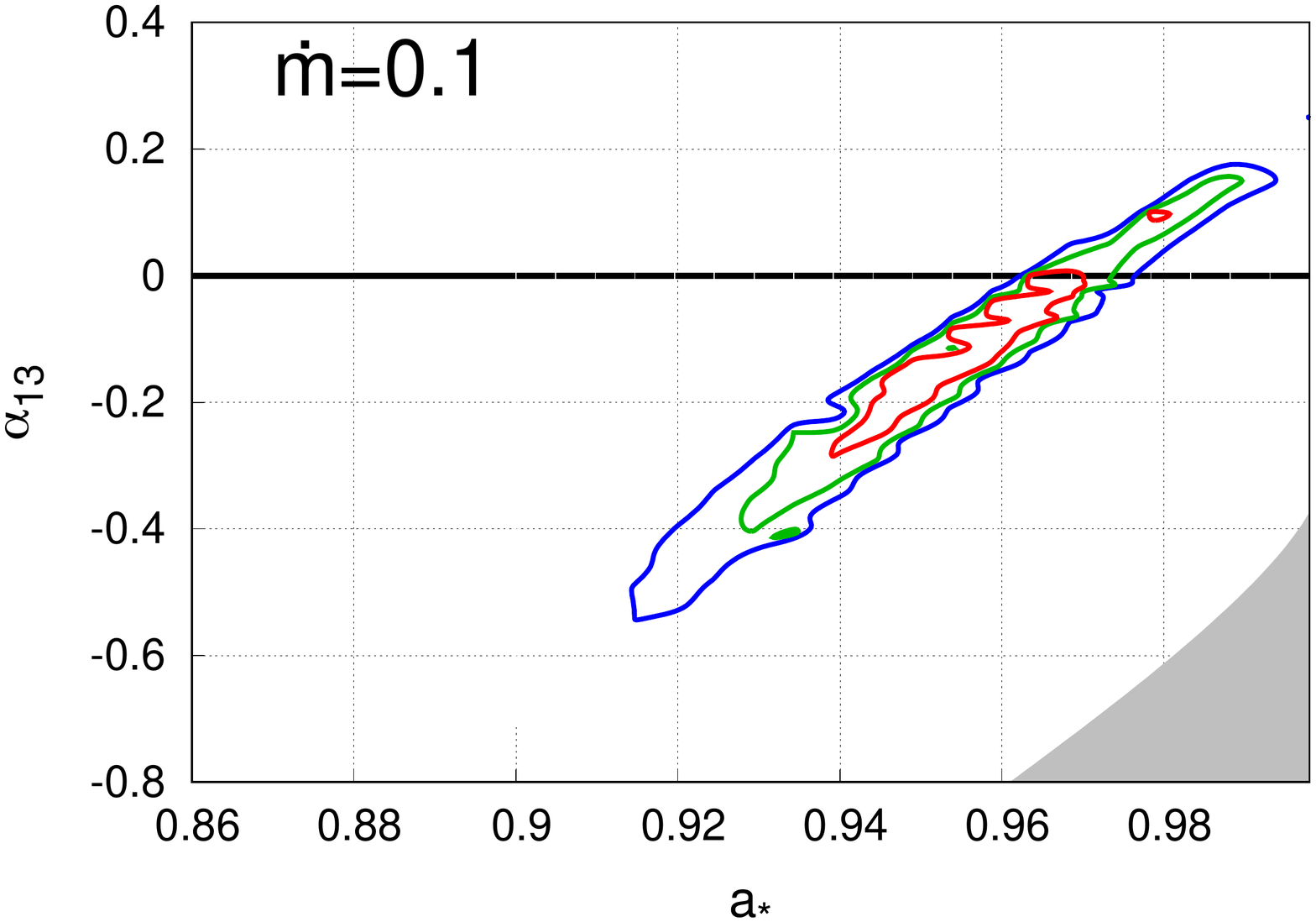} \\
\includegraphics[width=8.5cm,trim={0.5cm 1.5cm 0.5cm 0.5cm},clip]{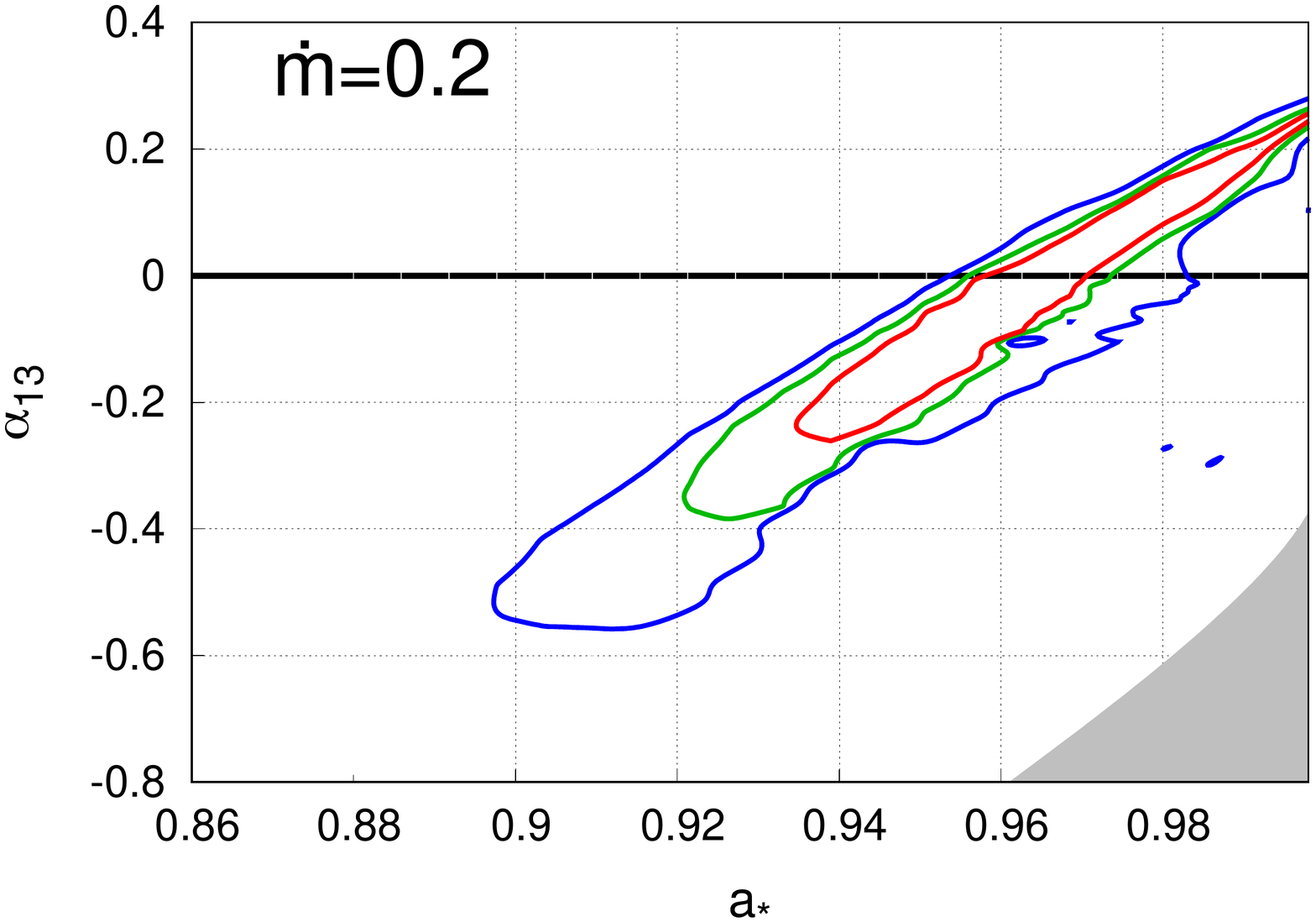}
\includegraphics[width=8.5cm,trim={0.5cm 1.5cm 0.5cm 0.5cm},clip]{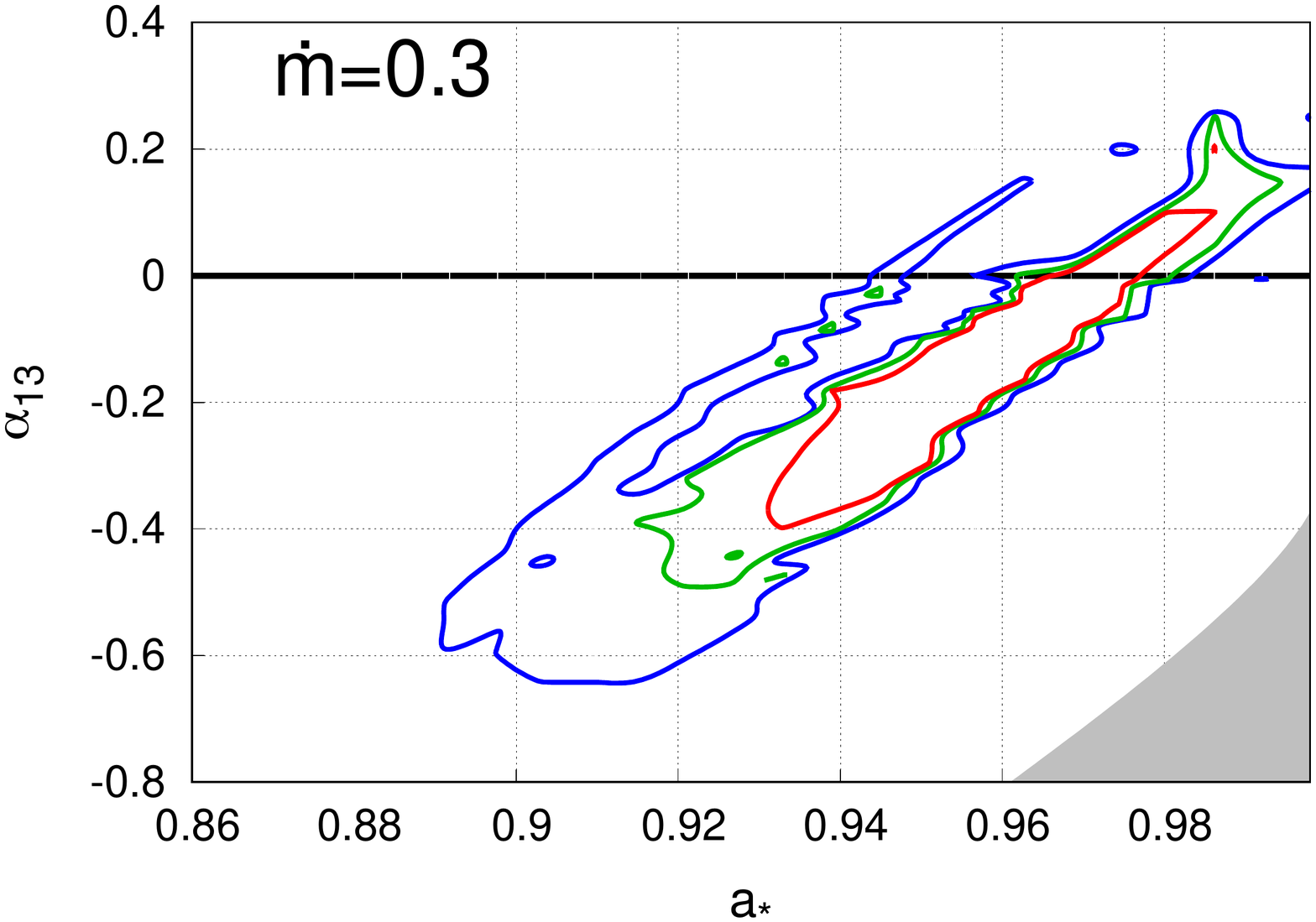}
\end{center}
\vspace{-0.5cm}
\caption{Constraints on the black hole spin parameter $a_*$ and the Johannsen deformation parameter $\alpha_{13}$ for the black hole in MCG--06--30--15 from Models~0-4 after marginalizing over all the free parameters. The red, green, and blue curves represent, respectively, the 68\%, 90\%, and 99\% confidence level curves for two relevant parameters. The gray region is ignored in our analysis because the spacetime is pathological there. The thick horizontal line at $\alpha_{13}=0$ marks the Kerr solution. \label{f-mcg}}
\end{figure*}

The results of the fits for Models~0-4 are all very similar. Fig.~\ref{f-mcgm} shows the best-fit models and the data to the best-fit model ratios for the four flux states for Model~0 ($\dot{m}=0$, infinitesimally thin disk). We do not show the same figures for Models~1-4 because we do not see any clear difference. Tab.~\ref{t-mcg-1}, Tab.~\ref{t-mcg-2}, and Tab.~\ref{t-mcg-3} show the summary of the best-fit values for Models~0-4. Again, the results are quite similar and the difference of $\chi^2$ among the different models is small. We also note that all our measurements are very similar to the estimates found in \citet{2019ApJ...875...56T} with an earlier version of {\tt relxill\_nk} and assuming an infinitesimally thin disk. A direct comparison with \citet{2014ApJ...787...83M} is not straightforward, because they use a different scheme to deal with the source variability, but still we recover consistent results \citep[the minor discrepancy on the black hole spin can be explained with the angle-averaged reflection model used in][see Section~\ref{s-dis}]{2014ApJ...787...83M}. The constraints on the black hole spin parameter $a_*$ and the Johannsen deformation parameter $\alpha_{13}$ for Models~0-4 and after marginalizing over all other free parameters are reported in Fig.~\ref{f-mcg}, where the red, green, and blue curves are, respectively, the 68\%, 90\%, and 99\% confidence level limits for two relevant parameters. The discussion of our spectral analysis of MCG--06--30--15 is postponed to Section~\ref{s-dis}.


\begin{figure}
\begin{center}
\includegraphics[width=8.5cm,trim={2.3cm 0.0cm 3.5cm 17.0cm},clip]{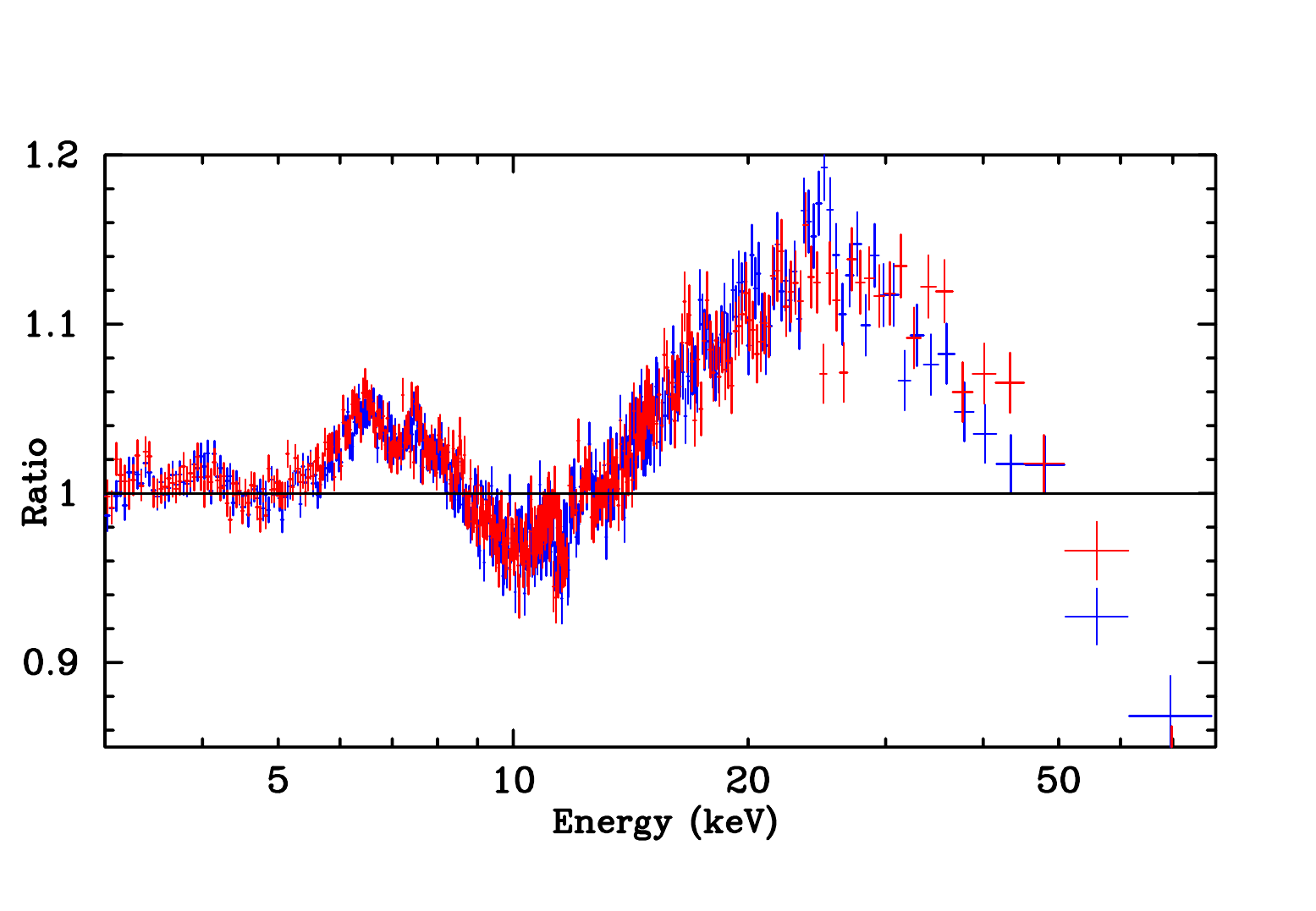}
\end{center}
\vspace{-0.7cm}
\caption{EXO~1846--031. Data to best-fit model ratio for an absorbed power-law. Blue crosses are for \textsl{NuSTAR}/FPMA and red crosses for \textsl{NuSTAR}/FPMB. 
\label{f-exor}}
\end{figure}

\begin{figure}
\begin{center}
\includegraphics[width=8.5cm,trim={2.3cm 0.0cm 2.5cm 17.0cm},clip]{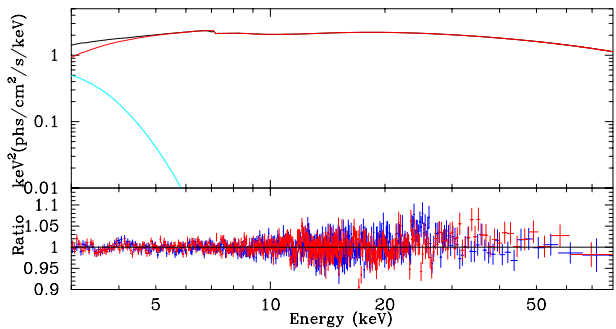}
\end{center}
\vspace{-0.7cm}
\caption{Best-fit model and data to the best-fit model ratio for EXO~1846--031 for Model~0 with an infinitesimally thin accretion disk. In the best-fit model plot, the total spectrum is in black, the {\tt relxill\_nk} component is in red, and the {\tt diskbb} component is in cyan. In the ratio plot, blue crosses are for \textsl{NuSTAR}/FPMA and red crosses for \textsl{NuSTAR}/FPMB. \label{f-exom}}
\end{figure}

\begin{table*}
\centering
{\renewcommand{\arraystretch}{1.3}
\begin{tabular}{lccccc}
\hline\hline
& Model~0 & Model~1 & Model~2 & Model~3 &  Model~4  \\
\hline\hline
{\tt tbabs} \\
$N_{\rm H}$ [$10^{22}$~cm$^{-2}$] & $10.8_{-0.8}^{+0.7}$ & $10.9_{-0.8}^{+0.6}$ & $10.9_{-0.8}^{+0.6}$ & $10.9_{-0.8}^{+0.7}$ & $10.8_{-0.6}^{+0.6}$ \\
\hline 
{\tt diskbb} \\
$kT_{\rm in}$ [keV] & $0.425_{-0.013}^{+0.009}$ & $0.426_{-0.011}^{+0.010}$ & $0.428_{-0.011}^{+0.010}$ & $0.430_{-0.011}^{+0.011}$ & $0.430_{-0.010}^{+0.011}$ \\
\hline
{\tt relxill\_nk} \\
$\dot{m}$ & $0^\star$ & $0.05^\star$ & $0.1^\star$ & $0.2^\star$ & $0.3^\star$ \\
$q_{\rm in}$ & $7.4_{-0.4}^{+0.4}$ & $7.38_{-1.02}^{+0.08}$ & $6.68_{-1.12}^{+0.14}$ & $5.9_{-0.8}^{+0.5}$ & $5.65_{-0.73}^{+0.10}$ \\
$q_{\rm out}$ & $0.1_{\rm -(P)}^{+0.8}$ & $0.05_{\rm -(P)}^{+0.46}$ & $0.05_{\rm -(P)}^{+0.51}$ & $0.05_{\rm -(P)}^{+0.9}$ & $0.05_{\rm -(P)}^{+0.18}$ \\
$R_{\rm br}$ [$r_{\rm g}$] & $8.2_{-4.6}^{+1.6}$ & $8.6_{-2.1}^{+1.0}$ & $10.4_{-1.4}^{+18.8}$ & $14_{-6}^{+4}$ & $14.9_{-4.4}^{+1.5}$ \\
$a_*$ & $0.998_{-0.002}^{}$ & $0.998_{-0.002}^{}$ & $0.998_{-0.002}^{}$ & $0.998_{-0.002}^{}$ & $0.998_{-0.002}^{}$ \\
$i$ [deg] & $73.3_{-0.6}^{+4.5}$ & $75.1_{-0.5}^{+2.8}$ & $75.1_{-0.5}^{+2.8}$ & $75_{-12}^{+3}$ & $76.3_{-0.8}^{+2.6}$ \\
$\Gamma$ & $2.001_{-0.010}^{+0.018}$ & $2.004_{-0.009}^{+0.018}$ & $2.003_{-0.008}^{+0.017}$ & $2.002_{-0.045}^{+0.018}$ & $2.001_{-0.006}^{+0.018}$ \\
$E_{\rm cut}$ [keV] & $198_{-9}^{+5}$ & $196_{-9}^{+6}$ & $196_{-9}^{+6}$ & $196_{-9}^{+7}$ & $196_{-9}^{+5}$ \\
$\log\xi$ [erg~cm~s$^{-1}$] & $3.46_{-0.05}^{+0.05}$ & $3.454_{-0.037}^{+0.023}$ & $3.46_{-0.04}^{+0.15}$ & $3.46_{-0.04}^{+0.06}$ & $3.463_{-0.03}^{+0.13}$ \\
$A_{\rm Fe}$ & $0.86_{-0.03}^{+0.08}$ & $0.85_{-0.07}^{+0.03}$ & $0.84_{-0.08}^{+0.03}$ & $0.84_{-0.04}^{+0.09}$ & $0.84_{-0.07}^{+0.03}$ \\
$R_{\rm f}$ & $1^\star$ & $1^\star$ & $1^\star$ & $1^\star$ & $1^\star$ \\
$\alpha_{13}$ & $0.00_{-0.08}^{+0.03}$ & $0.00_{-0.10}^{+0.02}$ & $0.00_{-0.11}^{+0.02}$ & $0.00_{-0.15}^{+0.02}$ & $0.00_{-0.18}^{+0.02}$ \\
\hline
{\tt gaussian} \\
$E_{\rm line}$ & $7.00_{-0.05}^{+0.06}$ & $7.00_{-0.05}^{+0.05}$ & $7.00_{-0.05}^{+0.05}$ & $7.01_{-0.05}^{+0.06}$ & $7.01_{-0.05}^{+0.05}$ \\
\hline
$\chi^2/{\rm dof}$ & 2751.86/2594 & 2753.49/2594 & 2754.98/2594 & 2757.26/2594 & 2758.18/2594 \\
& =1.06085 & =1.06148 & =1.06206 & =1.06294 & =1.06329 \\ 
\hline\hline
\end{tabular}
}
\caption{\rm Summary of the best-fit values for EXO~1846--031 for Models~0-4. The reported uncertainties correspond to the 90\% confidence level for one relevant parameter ($\Delta\chi^2 = 2.71$). $^\star$ indicates that the value of the parameter is frozen in the fit. (P) means that the 90\% confidence level reaches the upper boundary of the parameter. The best-fit of the black hole spin parameter is always stuck at its upper boundary $a_*^{\rm max} = 0.998$. \label{t-exo}}
\end{table*}

\begin{figure*}
\begin{center}
\includegraphics[width=8.5cm,trim={0.5cm 1.5cm 0.5cm 0.5cm},clip]{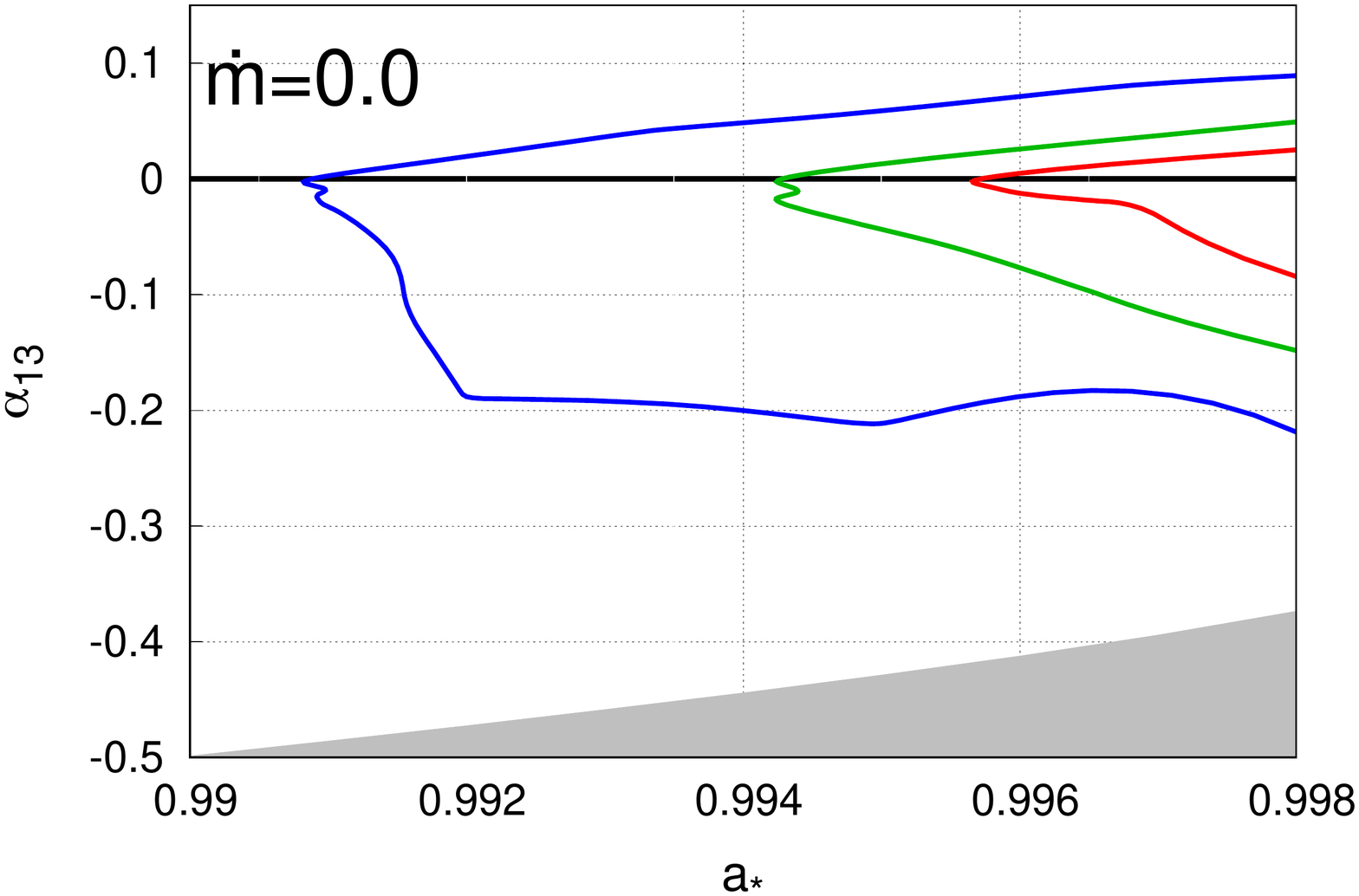} \\
\includegraphics[width=8.5cm,trim={0.5cm 1.5cm 0.5cm 0.5cm},clip]{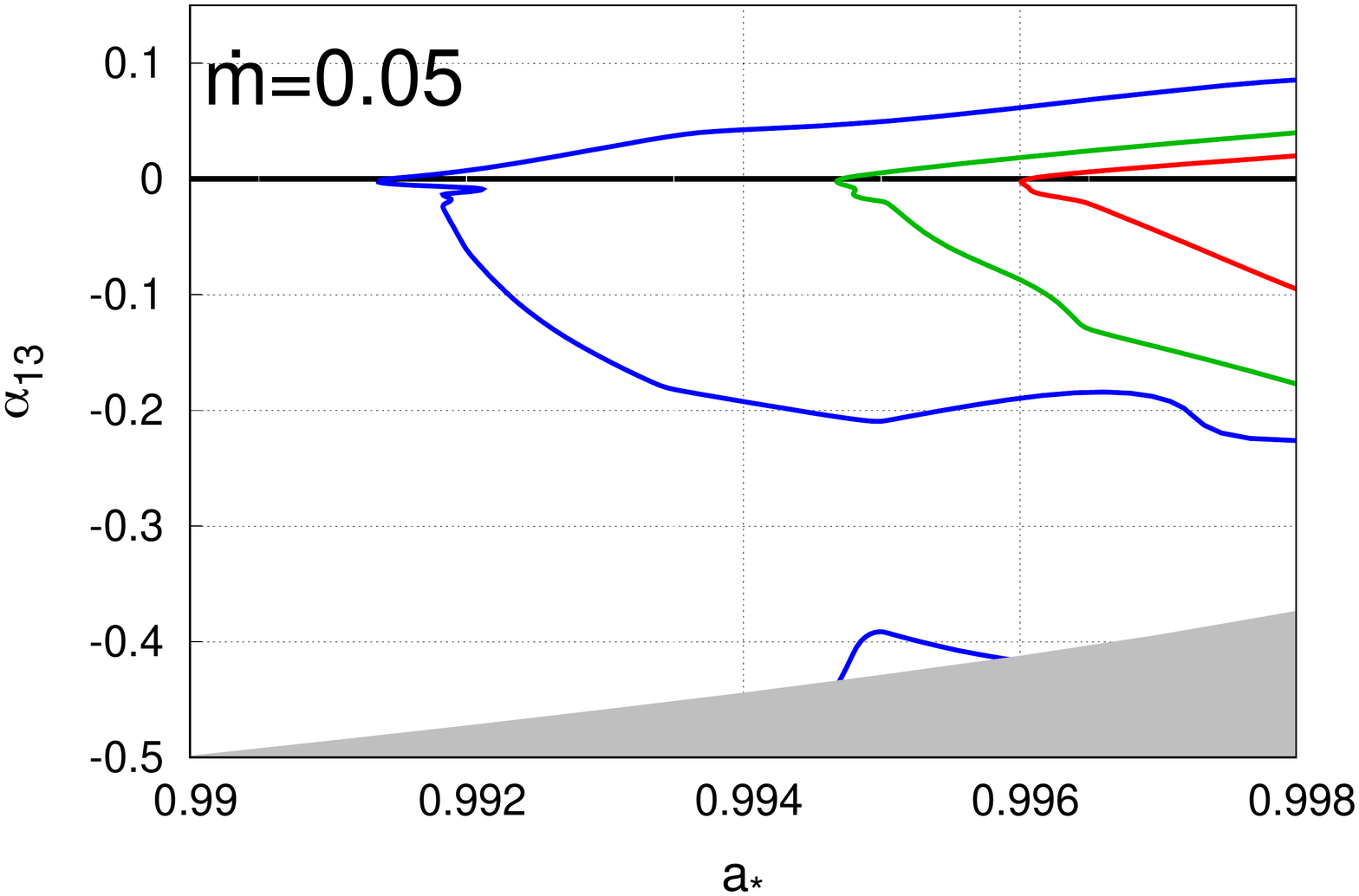}
\includegraphics[width=8.5cm,trim={0.5cm 1.5cm 0.5cm 0.5cm},clip]{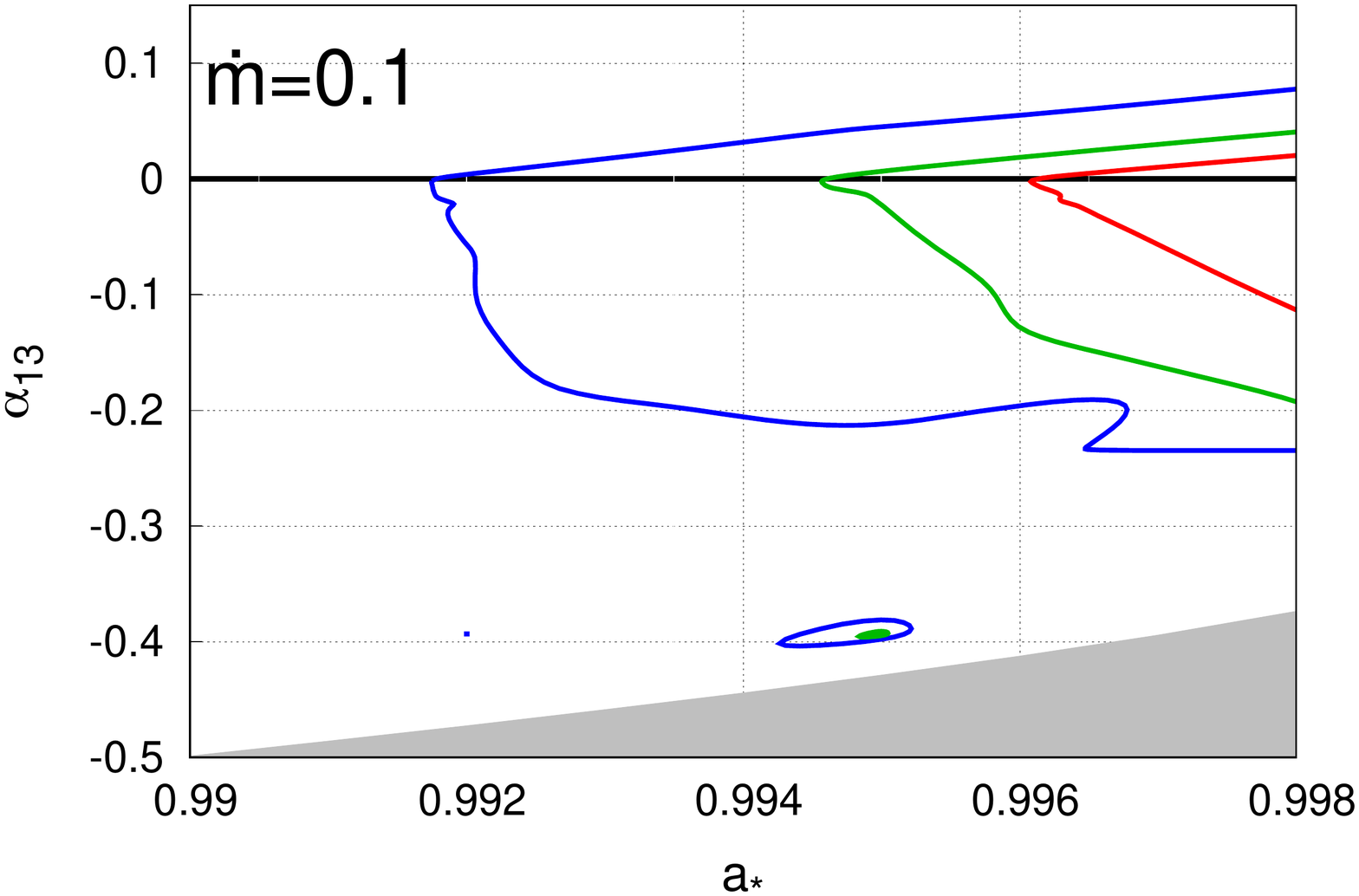} \\
\includegraphics[width=8.5cm,trim={0.5cm 1.5cm 0.5cm 0.5cm},clip]{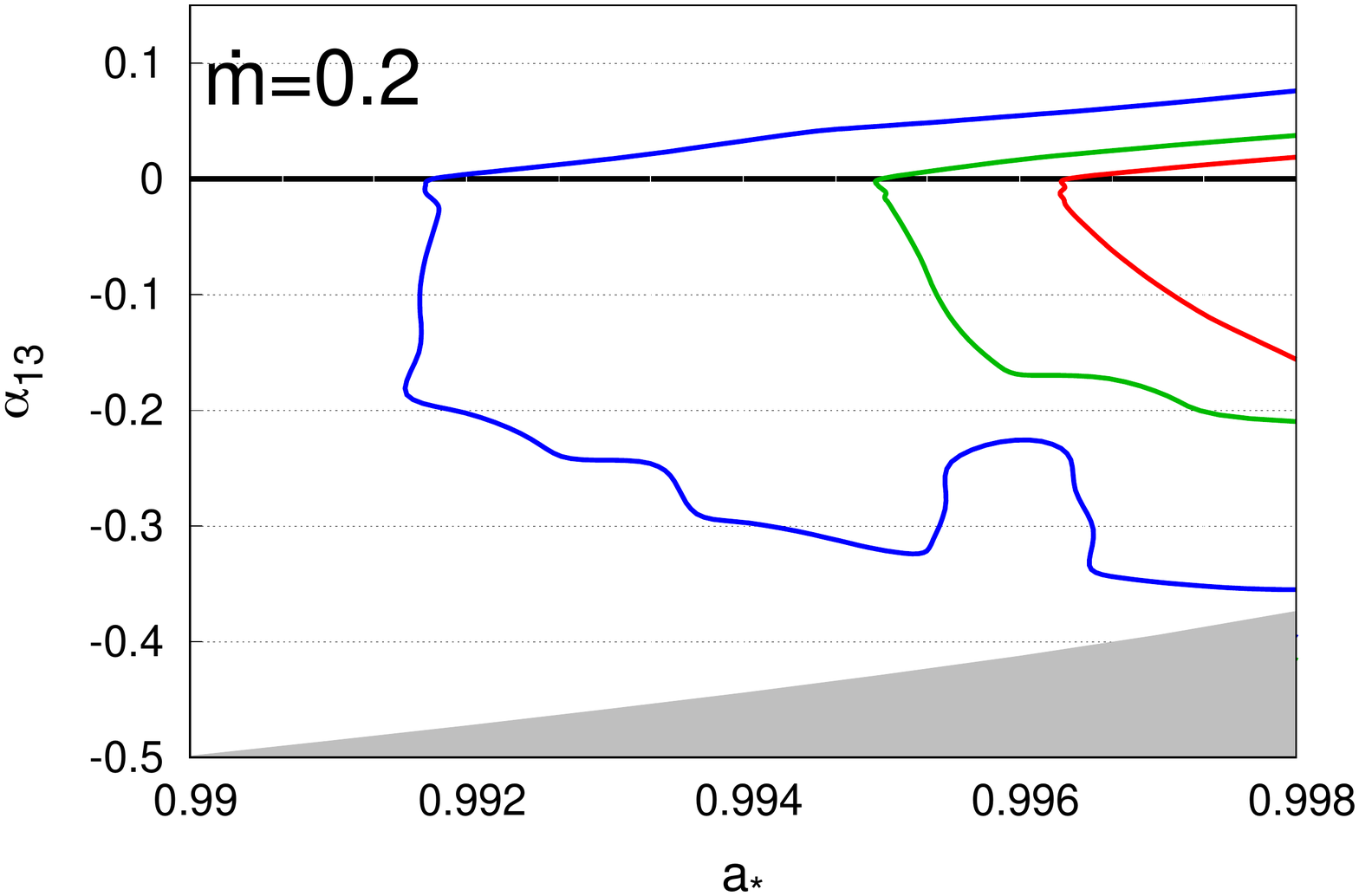}
\includegraphics[width=8.5cm,trim={0.5cm 1.5cm 0.5cm 0.5cm},clip]{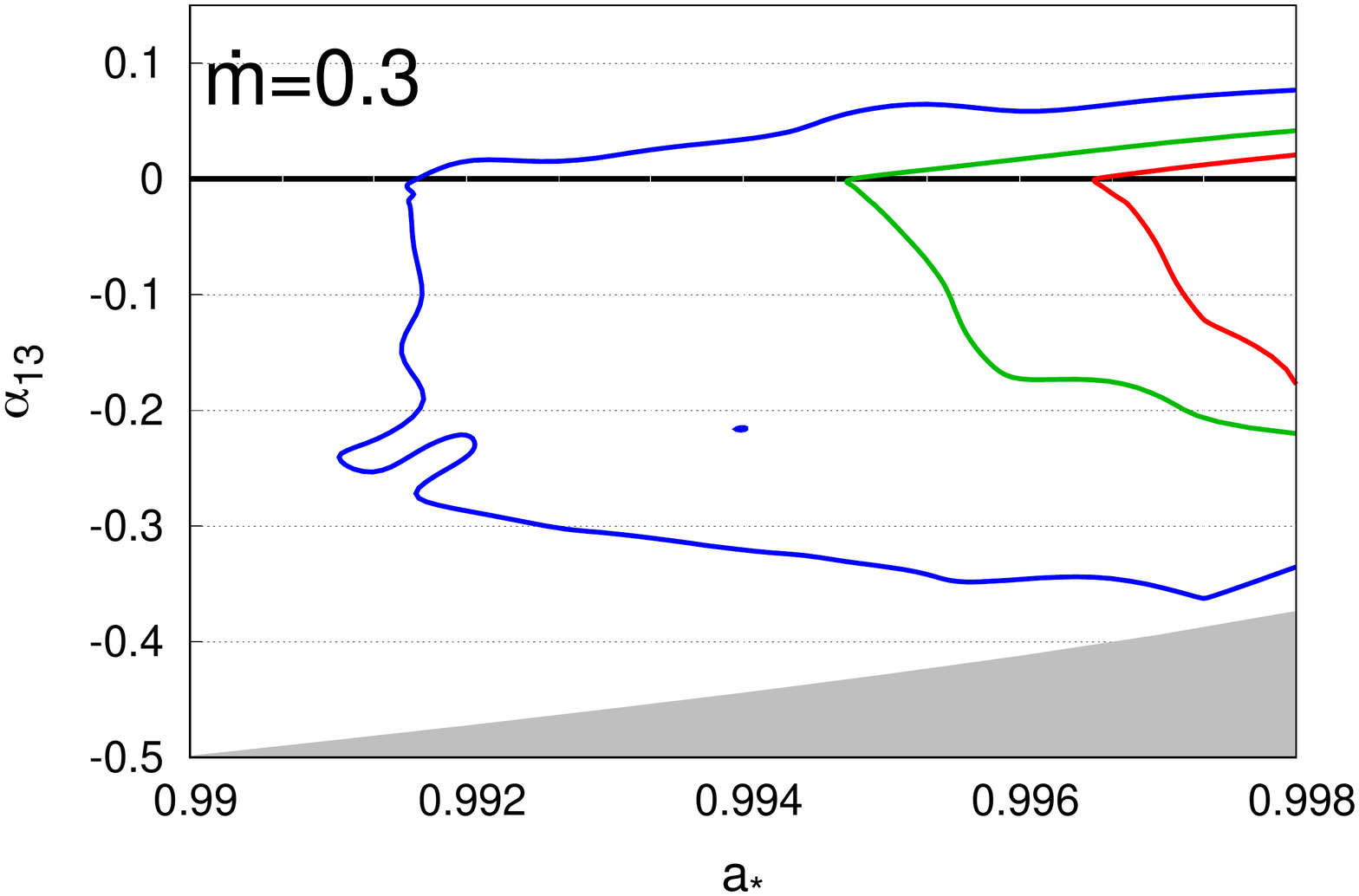}
\end{center}
\vspace{-0.5cm}
\caption{Constraints on the black hole spin parameter $a_*$ and the Johannsen deformation parameter $\alpha_{13}$ for the black hole in EXO~1846--031 from Models~0-4 after marginalizing over all the free parameters. The red, green, and blue curves represent, respectively, the 68\%, 90\%, and 99\% confidence level curves for two relevant parameters. The gray region is ignored in our analysis because the spacetime is pathological there. The thick horizontal line at $\alpha_{13}=0$ marks the Kerr solution. \label{f-exo}}
\vspace{0.4cm}
\end{figure*}

\section{EXO~1846--031} \label{s-exo}

EXO~1846--031 was first detected by \textsl{EXOSAT} on 1985 April 3~\citep{1985IAUC.4051....1P} and only later it was identified as a low mass X-ray binary~\citep{1993A&A...279..179P}. A second outburst of this source was detected in 1994 by \textsl{CGRO}/BATSE~\citep{1994IAUC.6096....1Z}. After about 25~years of quiescence, \textsl{MAXI} observed a third outburst of EXO~1846--031 in July 2019~\citep{2019ATel12968....1N}. Following the \textsl{MAXI} detection, the source was observed with other instruments. In this section, we use the \textsl{NuSTAR} data, first analyzed in \citet{2020ApJ...900...78D}.

\subsection{Data reduction}

\textsl{NuSTAR} observed EXO~1846--031 on 2019 August 3 for about 22~ks (see Tab.~\ref{t-obs}). Unlike the observation of MCG--06--30--15 analyzed in the previous section, the \textsl{NuSTAR} light curve of EXO~1846--031 shows minimal variability in the count rate and therefore we can directly use the time-averaged spectrum in our spectral analysis.

The raw data are reduced to cleaned event files using the NUPIPELINE routine of the HEASOFT package with CALDB v20200912. For the source spectra, we take a circular region with the radius of 180~arcsec centered around the source. We extract a background region of the same size as far as possible from the 
source on the same detector to avoid any source contamination. The ancillary and response files are generated through the NUPRODUCT routine. The source spectra are rebinned to have a minimum of 30~counts per bin in order to use $\chi^2$-statistics.

\subsection{Spectral analysis}

We fit the FPMA and FPMB spectra together in the energy range 3-80~keV. To start, we fit the data with an absorbed power-law and the data to the best-fit model ratio is shown in Fig.~\ref{f-exor}. We clearly see a broad iron line peaked around 7~keV and a Compton hump peaked around 30~keV. The fact that the iron line is so broad already suggests that the inner edge of the accretion disk extend to a region very close to the black hole event horizon.

To fit these reflection features, we use {\tt relxill\_nk} with the spacetime described by the Johannsen metric with non-vanishing deformation parameter $\alpha_{13}$ and an accretion disk of finite thickness. It is the same version of {\tt relxill\_nk} as that used in the previous section for MCG--06--30--15. The inner edge is set at the ISCO radius and the outer edge is left at the default value of 400~$r_{\rm g}$. The emissivity profile is modeled with a broken power-law and we have three free parameter: the inner emissivity index $q_{\rm in}$, the outer emissivity index $q_{\rm out}$, and the breaking radius $R_{\rm br}$. The reflection fraction cannot be constrained well and therefore it is frozen to 1, as in \citet{2020ApJ...900...78D}. As in the previous section for the spectral analysis of MCG--06--30--15, the mass accretion rate, $\dot{m}$, is frozen to 0 (Model~0), 0.05 (Model~1), 0.1 (Model~2), 0.2 (Model~3), and 0.3 (Model~4).

The fit of the model {\tt tbabs$\times$relxill\_nk} still presents some residuals at low energies, which we remove by adding a continuum component from the accretion disk with {\tt diskbb}. The latter is a Newtonian model for an infinitesimally thin disk, but the residual is small and it is not necessary to employ a more sophisticated model. We still see an absorption feature around 7~keV, which can be interpreted as absorption by material in the disk wind at a relatively high inclination angle. Similar features are observed in other sources, for instance 4U~1630--472~\citep{2014ApJ...784L...2K}. We model this absorption feature with a Gaussian.

The fluxes of FPMA and FPMB clearly appears different at low energies. As discussed in \citet{2020arXiv200500569M}, it is likely an instrumental issue in FPMA. Here we follow the same procedure as in \citet{2020ApJ...900...78D}. First, we fit the FPMA and FPMB spectra together with a free constant cross-calibration and ignoring the 3-7~keV energy band. We freeze the cross-calibration constant to the value found by our fit and we include the data in the 3-7~keV energy band. At this point, we add the multiplicative table described in \citet{2020arXiv200500569M}. We fix the MLI fraction in the FPMB spectrum to 1 and we allow the MLI faction in the FPMA spectrum to vary. In the end, our total model is 

\vspace{0.2cm}

\noindent {\tt tbabs$\times$nuMLI$\times$(relxill\_nk+diskbb+guassian)} .

\vspace{0.2cm}

Tab.~\ref{t-exo} shows the best-fit values for Models~0-4. The best-fit model and the data to the best-fit model ratio for Model~0 are shown in Fig.~\ref{f-exom}. Here we do not report the same figures for Models~1-4 but they are all very similar to Fig.~\ref{f-exom}. Fig.~\ref{f-exo} shows the constraints on the black hole spin parameter $a_*$ and the Johannsen deformation parameter $\alpha_{13}$ for Models~0-4. The red, green, and blue curves represent, respectively, the 68\%, 90\%, and 99\% confidence level limits for two relevant parameters. The discussion of our results is in the next section.


\section{Discussion and conclusions} \label{s-dis}

Within our program of development of the reflection model {\tt relxill\_nk}, in \citet{2020ApJ...899...80A} we presented a version in which the accretion disk has a finite thickness, following the implementation first proposed in \citet{2018ApJ...855..120T}. In the present work,  we have used that new version of {\tt relxill\_nk} to fit the high-quality spectra of two sources in order to evaluate the impact of the disk thickness on the estimate of the model parameters, and in particular on the measurement of the black hole spin parameter $a_*$ and the Johannsen deformation parameter $\alpha_{13}$. For this study, we have considered the 2013 simultaneous observations of \textsl{NuSTAR} and \textsl{XMM-Newton} of MCG--06--30--15 and the 2019 \textsl{NuSTAR} observation of EXO~1846--031. Previous analyses of these spectra had reported very precise constraints on $a_*$ and $\alpha_{13}$, so that changes in the theoretical model can potentially lead to a different measurement of these parameters. Moreover, these two sources are different, as MCG--06--30--15 hosts a supermassive black hole observed from a low viewing angle and EXO~1846--031 hosts a stellar-mass black hole observed from a high viewing angle. Previous analyses had also reported a different disk's emissivity profiles for the two sources, suggesting a different coronal geometry, with MCG--06--30--15 with a high value of the inner emissivity index and a value of the outer emissivity index consistent with 3 and EXO~1846--031 with a high value of the inner emissivity index and a value of the outer emissivity index close to 0. The spectrum of every source has been fit with five different Eddington-scaled mass accretion rates: 0\% (corresponding to the standard model with an infinitesimally thin disk), 5\%, 10\%, 20\%, and 30\%.

For MCG--06--30--15, the five models provide very similar results, as we can see in Tab.~\ref{t-mcg-1}, Tab.~\ref{t-mcg-2}, and Tab.~\ref{t-mcg-3}. These results are also very similar to those found in our previous analysis with an older version of {\tt relxill\_nk} and an infinitesimally thin accretion disk \citep{2019ApJ...875...56T}, and are consistent with those reported in \citet{2014ApJ...787...83M}, even if a direct comparison is not possible for the parameters that vary among flux states because \citet{2014ApJ...787...83M} use a different scheme. We note that here and in \citet{2019ApJ...875...56T} we find a somewhat higher value of the black hole spin parameter than that inferred in \citet{2014ApJ...787...83M}. Such a discrepancy is mainly due to the difference between the angle-resolved reflection model {\tt relxill}/{\tt relxill\_nk} and the angle-averaged model {\tt relconv$\times$xillver} employed in \citet{2014ApJ...787...83M} \citep[more details can be found in][]{2020MNRAS.498.3565T}. We also note that the $\chi^2$ of Models~0-4 are very similar. Model~3 ($\dot{m}=0.2$) has the lowest $\chi^2$ (3027.47) and Model~1 ($\dot{m}=0.05$) has the highest one (3028.67): the difference is only $\Delta\chi^2 = 1.20$. The weak impact of the disk thickness on the reflection spectrum of a high spin/high radiative efficiency source observed from a small viewing angle is already clear in the plot in the top-right corner in Fig.~\ref{fig:eemod}. Even if our current version of {\tt relxill\_nk} cannot have $\dot{m}$ free, we can easily argue that we could not measure it in MCG--06--30--15, in the sense that any attempt to infer its value from the reflection spectrum would give an unconstrained parameter.

The analysis of the \textsl{NuSTAR} spectrum of EXO~1846--031 leads to similar, even if not identical, conclusions. Tab.~\ref{t-exo} shows the best-fit values of the model parameters. As can be also seen from Fig.~\ref{f-exo}, the thickness of the disk does not seem to have any significant impact on the estimates of the black hole spin parameter $a_*$ and the Johannsen deformation parameter $\alpha_{13}$: the final constraints are extremely similar for Models~0-4. This is somewhat more remarkable, because the source is observed with a very high inclination angle, so a region of the very inner part of the accretion disk may be obscured. As we can see from the bottom-right panel in Fig.~\ref{fig:eemod}, spectra for different $\dot{m}$ are not very similar as in the case of a low viewing angle. Note, however, that the extremely high estimate of the spin parameter of the source indicates a very high radiative efficiency and, therefore, the thickness of the disk should be quite modest; see Fig.~\ref{fig:disk}. For EXO~1846--031, we find a more pronounced difference among the $\chi^2$ of models with different mass accretion rates. The lowest $\chi^2$ is found for the infinitesimally thin disk model (Model~0, 2751.86) and $\chi^2$ monotonically increases as $\dot{m}$ increases, reaching $\chi^2 = 2758.18$ for Model~4 with $\dot{m}=0.3$. Such a trend, which cannot be seen in the analysis of MCG--06--30--15, where all fits are very similar because of the low viewing angle, should be investigated better in the future with other spectra: it seems like the model with an infinitesimally thin disk fits the data better than the models with a disk of finite thickness and thus the disk structure employed in {\tt relxill\_nk} is unsuitable to describe real accretion disks around black holes, or at least the accretion disk around the black hole in EXO~1846--031 during the \textsl{NuSTAR} observation. We note that the mass and the distance are poorly constrained for EXO~1846--031. \citet{2020ApJ...900...78D} estimate that the Eddington-scaled disk luminosity in this observation is in the range 0.06 to 0.25 assuming the black hole mass $M = 9 \pm 5~M_\odot$ (from the continuum-fitting method with 1-$\sigma$ error) and the distance $D = 7$~kpc. If we calculate the 0.1-100~keV unabsorbed flux for our best-fit model, we recover a similar result. However, the uncertainty is so large that we have no indication of which value of $\dot{m}$ we should use to fit the reflection spectrum.

The thickness of the disk seems to have the effect of decreasing the estimate of the inner emissivity index $q_{\rm out}$ and increasing the measurement of the breaking radius $R_{\rm br}$. The estimate of the other model parameters does not show any variation with different values of the mass accretion rate $\dot{m}$. The iron abundance, $A_{\rm Fe}$, is estimated to be below the solar abundance, regardless of the value of $\dot{m}$, and this might be caused by the fact that our model does not include the returning radiation in the calculation of the reflection spectrum and the viewing angle of the source is high \citep[see the discussion in][]{2020arXiv200615838R}.

The impact of the thick disk model on the estimate of $q_{\rm out}$ and $R_{\rm br}$ indicates the presence of a correlation between the thickness of the disk and the emissivity profile (and, consequently, the coronal geometry). If we freeze $q_{\rm in}$, $q_{\rm out}$, and $R_{\rm br}$ to the best-fit values found with the infinitesimally thin disk model and we refit the data with Models~1-4, we find that only the estimate of the inclination angle of the disk changes, see Tab.~\ref{t-exo-bis}, while all other parameters, including the black hole spin and the deformation parameter $\alpha_{13}$, do not show any appreciable difference. Such a result support the claim of the robustness of the estimates of the black hole spin parameter and of the deformation parameter $\alpha_{13}$, at least in the high radiative efficiency regime.

\begin{table*}
\centering
{\renewcommand{\arraystretch}{1.3}
\begin{tabular}{lcccc}
\hline\hline
& Model~1 & Model~2 & Model~3 &  Model~4  \\
\hline\hline
{\tt relxill\_nk} \\
$\dot{m}$ & $0.05^\star$ & $0.1^\star$ & $0.2^\star$ & $0.3^\star$ \\
$\iota$ [deg] & $75.2_{-0.4}^{+0.4}$ & $77.2_{-0.4}^{+0.5}$ & $80.2_{-0.4}^{+0.4}$ & $82.6_{-0.9}^{+0.6}$ \\
\hline
$\chi^2/{\rm dof}$ & 2753.66/2597 & 2756.97/2597 & 2765.05/2597 & 2767.69/2597 \\
& =1.06032 & =1.06160 & =1.06471 & =1.06573 \\ 
\hline\hline
\end{tabular}
}
\caption{\rm Best-fit values of the inclination angle of the disk, $\iota$, and $\chi^2/{\rm dof}$ for EXO~1846--031 for Models~1-4 when we fit the data freezing $q_{\rm in}$, $q_{\rm out}$, and $R_{\rm br}$ to the best-fit values found in the fit of Model~0. The estimates of the other parameters do not show any appreciable difference. \label{t-exo-bis}}
\end{table*}

In conclusion, our analysis of MCG--06--30--15 and EXO~1846--031 suggests that the thickness of the disk has quite a modest impact on the fit of the reflection features in general, and on the estimate of the black hole spin parameter and the Johannsen deformation parameter in particular. We want to remark that we are still assuming a Novikov-Thorne accretion disk with inner edge at the ISCO radius. The accretion disk is thus geometrically thin, and for this reason we limit the model to the range $\dot{m} = 0$-0.3; we simply have a thin disk of finite thickness and the thickness increases as the mass accretion rate increases. The scenario studied in \citet{2020ApJ...895...61R} and \citet{2020MNRAS.491..417R} was different: in those papers we simulated the reflection spectra of geometrically thick accretion disks and we found we could easily get very precise but inaccurate estimates of the black hole spin parameter and the Johannsen deformation parameter, which is the opposite conclusion of the present work. We would like also to point out that our finding is not inconsistent with the conclusions of \citet{2018ApJ...855..120T}, where the authors fit some simulated spectra and find that the thickness of the disk can lead to inaccurate estimates of the black hole spin parameter and the disk's inclination angle. In \citet{2018ApJ...855..120T}, the simulations are done assuming a black hole spin $a_* = 0.9$ (so with a lower radiative efficiency with respect to our sources) and a lamppost coronal geometry. If the height of the lamppost is low, the intensity profile is very different from the case of an infinitesimally thin disk, where at large radii the emissivity should scale as $1/r^3$. Here we fit the data of two sources in which the geometry of the corona is unknown, but presumably different from the lamppost set-up (which, assuming an infinitesimally thin disk, provides a fit worse than the broken-power law model). The disk's intensity profile, which depends on the coronal geometry, seems thus to be crucial to determine whether the thickness of the disk has or does not have an important impact on the estimate of the model parameters. This is consistent with the conclusions found in \citet{2020ApJ...895...61R} and \citet{2020MNRAS.491..417R} for geometrically thick disks, where the estimate of the black hole spin parameter is strongly determined by the disk's emissivity profile when we use an incorrect theoretical model.


\vspace{0.5cm}

{\bf Acknowledgments --}
This work was supported by the Innovation Program of the Shanghai Municipal Education Commission, Grant No.~2019-01-07-00-07-E00035, the National Natural Science Foundation of China (NSFC), Grant No.~11973019, and Fudan University, Grant No.~JIH1512604. 
A.T., C.B., and H.L. are members of the International Team~458 at the International Space Science Institute (ISSI), Bern, Switzerland, and acknowledge support from ISSI during the meetings in Bern.


\end{document}